\documentclass[superscriptaddress,aps,prd,nofootinbib,twocolumn,showpacs]{revtex4-1}
\usepackage{amsfonts}
\usepackage{amsmath}
\usepackage{amssymb}
\usepackage{graphicx}
\usepackage{mathrsfs}
\usepackage{hhline,color}
\usepackage{pifont,ulem}
\usepackage{lmodern}
\usepackage{epsfig}
\usepackage{enumitem} 
\usepackage{ulem}
\usepackage[utf8]{inputenc}

\newcommand{\pc}[1]{\ensuremath{\left(#1\right)}}

\newcommand{\ev}[1]{\ensuremath{\left\langle #1\right\rangle}}

\def\beq{\begin{equation}}
\def\eeq{\end{equation}}
\def\beqa{\begin{eqnarray}}
\def\eeqa{\end{eqnarray}}
\def\ban{\begin{eqnarray*}}
\def\ean{\end{eqnarray*}}
\def\bi{\begin{itemize}}
\def\ei{\end{itemize}}

\newcommand{\Z}{\mathbb{Z}}

\begin{document}

\title{Influence of the inverse magnetic catalysis and the vector 
interaction in the location of the critical end point}

\author{Pedro Costa}
\email{pcosta@teor.fis.uc.pt}
\affiliation{CFisUC, Department of Physics,
University of Coimbra, P-3004 - 516  Coimbra, Portugal}
\author{ M\'arcio Ferreira}
\email{mferreira@teor.fis.uc.pt}
\affiliation{CFisUC, Department of Physics,
University of Coimbra, P-3004 - 516  Coimbra, Portugal}
\author{D\'ebora P. Menezes}
\email{debora.p.m@ufsc.br}
\affiliation{Departamento de F\'{\i}sica, Universidade Federal de Santa Catarina, 
Florian\'{o}polis, SC, CP.476, CEP 88.040-900, Brazil}
\author{Jo\~ao Moreira}
\email{jmoreira@teor.fis.uc.pt}
\affiliation{CFisUC, Department of Physics,
University of Coimbra, P-3004 - 516  Coimbra, Portugal}
\author{Constan\c ca Provid\^encia}
\email{cp@teor.fis.uc.pt}
\affiliation{CFisUC, Department of Physics,
University of Coimbra, P-3004 - 516  Coimbra, Portugal}

\date{\today}

\begin{abstract}
The effect of a strong magnetic field on the location of the critical end
point (CEP) in the QCD phase diagram is discussed under different scenarios. 
In particular, we consider the contribution of the vector interaction and take 
into account the inverse magnetic catalysis obtained in lattice QCD calculations 
at zero chemical potential.  The discussion is realized within the (2+1)
Polyakov--Nambu--Jona-Lasinio model. It is shown that the vector interaction and
the magnetic field have opposite competing effects, and that the winning 
effect depends strongly on the intensity of the magnetic field. 
The inverse magnetic catalysis at zero chemical potential has two distinct 
effects for magnetic fields above $\gtrsim 0.3$ GeV$^2$: it shifts the CEP to 
lower chemical potentials, hinders the increase of the  CEP temperature and 
prevents a too large increase of the baryonic density at the CEP. 
For fields $eB<0.1$ GeV$^2$ the competing effects between the vector contribution
and the magnetic field can move the CEP to regions of temperature and density 
in the phase diagram that could be more easily accessible to experiments.
\end{abstract}


\maketitle

\section{Introduction}

The existence and the location of the critical end point (CEP) is a very timely 
topic for theoretical studies based on QCD with direct implications on the 
nature of the phase transition between the hadron gas and the quark gluon plasma.
During the last decade, the investigation of the QCD phase diagram and the possible 
existence of the CEP has been attracting the attention of the physics community:
by using lattice QCD (LQCD) simulations \cite{Forcrand:2003hx,Fodor:2004nz}, 
Dyson-Schwinger equations \cite{Fischer:2014ata} and effective models namely 
the Nambu--Jona-Lasinio (NJL) model \cite{Costa:2008yh}, its recent extension 
up to eight quark terms (including explicit chiral symmetry breaking 
interactions) \cite{Moreira:2014qna} and the Polyakov--Nambu--Jona-Lasinio 
(PNJL) model \cite{Costa:2008gr}, there has been an effort to understand the 
nature of the phase transition and the existence of the CEP.

From the experimental point of view the existence/location of the CEP is the 
major goal of several programs namely the Beam Energy Scan (BES-I) program at 
RHIC which has been ongoing since 2010 looking for the experimental signatures 
of a first-order phase transition and the CEP by colliding Au ions at several 
energies \cite{Abelev:2009bw}. 
Recently, the results of the moments of net-charge multiplicity distributions
were presented by STAR Collaboration \cite{Adamczyk:2014fia}. These measurements 
can provide relevant information on the freeze-out conditions and can help to 
clarify the existence of the CEP. 
However, future measurements with high statistics data will be needed for a 
precise determination of the freeze-out conditions and to make
definitive conclusions regarding the CEP \cite{Adamczyk:2014fia}.
Also the dynamics associated with heavy-ion collisions, such as finite 
correlation length and freeze-out effects, should be considered in QCD 
calculations before definitive conclusions about the CEP can be made
\cite{Adamczyk:2013dal}.
If there is a CEP with baryonic chemical potential, $\mu_B$,  lower than 400 MeV, 
it is expected that the upcoming BES-II program can provide data on fluctuation 
and flow observables which should yield quantitative evidence for its presence. 
Otherwise, late in the decade, the FAIR facility at GSI and NICA at JINR will 
extend the search of the CEP to even higher $\mu_B$ (for a review on the 
experimental search of the CEP see \cite{Akiba:2015jwa}).
Also the NA61/SHINE program at the CERN SPS aims the search of the CEP, and to
investigate  the properties of the onset of deconfinement through spectra, 
fluctuations and correlations analysis in light and heavy ion collisions
\cite{Gazdzicki:2011fx}.

There are several aspects that can influence the location of the CEP like the
strangeness or isospin content of the in-medium or the presence of an external 
magnetic field \cite{Costa:2013zca}. Indeed, in Ref. \cite {Avancini:2012ee}, 
within the NJL model, it was verified that the size of the 
first order segment of the transition line expands with increasing $B$ in such 
a way that the CEP becomes located at higher temperature and smaller chemical 
potential values. This was also verified by using the Ginzburg-Landau effective 
action formalism with the renormalized quark-meson model \cite{Ruggieri:2014bqa}. 
The influence of strong external magnetic fields on the structure of the 
QCD phase diagram is also very important because it has relevant consequences 
on measurements in heavy-ion collisions at very high energies \cite{HIC}.

At finite temperature, several LQCD studies have been performed to address the 
influence of the magnetic field over the deconfinement and chiral
transition temperatures 
\cite{baliJHEP2012,bali2012PRD,endrodi2013,Ilgenfritz:2013ara,D'Elia:2010nq,
Bornyakov:2013eya}. For a review in recent advances in the understanding of the 
phase diagram in the presence of strong magnetic fields at zero quark chemical 
potentials see \cite{Andersen:2013swa}.

The inclusion of a magnetic field in the Lagrangian density of the NJL model and 
of the PNJL model allows describing the magnetic  catalysis (MC) effect, i.e., 
the enhancement of the quark condensate due to the magnetic field 
\cite{Fu:2013ica,Ferreira:2013tba,Ferreira:2013oda,Ferreira:2014exa}, but does 
not describe the suppression of the quark condensate found in LQCD calculations 
at finite temperature and zero chemical potential which is due to the strong 
screening effect of the gluon interactions, the so-called inverse magnetic 
catalysis (IMC) \cite{baliJHEP2012,bali2012PRD,endrodi2013}.
In order to deal with this discrepancy, it was proposed that the model coupling, 
$G_s$, can be seen as proportional to the running coupling, $\alpha_s$, and 
consequently, a decreasing function of the magnetic field strength allowing one 
to include its effects ($\alpha_s(eB)$). 
In fact, in the region of low momenta the strong screening effect of 
the gluon interactions weakens the interaction which leads to a decrease of the 
scalar coupling with the intensity of the magnetic field \cite{Miransky:2002rp}.
By using the SU(2) NJL model \cite{Farias:2014eca} and the SU(3) NJL/PNJL models 
\cite{Ferreira:2014kpa} two ansatz were proposed that allow for the IMC.

Other mechanisms that lead to the IMC can be found in the literature 
\cite{Ferreira:2013tba,Chao:2013qpa,Fukushima:2012kc,Ayala:2014iba,
Braun:2014fua,Mueller:2015fka}, together with a model-independent physical 
explanation for the IMC \cite{Preis:2010cq} while a review with analytical 
results for the NJL can be found in Ref. \cite {Preis:2012fh}. 

Another aspect that is relevant for the location of the CEP is the presence of 
the vector interaction which acts in the opposite way of the magnetic field
\cite{Denke:2013gha,Garcia:2013eaa}. 
Indeed, it is known that increasing the repulsive interaction strength 
in the quark matter phase diagram leads to a shrinking of the first-order 
transition region as the baryonic chemical potential increases (the CEP moves 
to larger $\mu_B$ and lower temperature $T$ \cite{Fukushima:2008wg}).
Furthermore, the increase of $G_V$ can change the structure of the
phase diagram by decreasing the possible quarkyonic phase \cite{Dutra:2013lya}. 
It is also important to note that, as pointed out in 
\cite{Fukushima:2008wg}, there is no constraint for the choice of the coupling 
$G_V$ at finite density; if we see $G_V$ as induced in dense quark matter, 
the choice of $G_V$ is not related with the vector meson properties in the vacuum 
but it can be related with in-medium modifications \cite{Hatsuda:1994pi}. 

The presence of a vector interaction also becomes relevant in reproducing some 
experimental results  \cite{Bernard:1993wf} or compact star
observations (see for instance \cite{sedrakian11}), and so should also be taken 
into account in the computation of the equation of state (EOS) for magnetized 
quark matter \cite{Menezes:2014aka,Chu:2014pja}. 
A step toward this type of investigation has recently been taken in Ref. 
\cite{Denke:2013gha}  where  two flavor magnetized quark matter in the presence 
of a repulsive vector coupling, described by the NJL model, has been considered.

In the present work we investigate the influence of the IMC and the vector 
interaction in the location of the CEP in magnetized matter using the
(2+1) PNJL model. 
After the presentation of the model in Sec. \ref{sec:model} several scenarios of 
interest will be explored: the influences of an external magnetic field and of 
the vector interaction in the location of the CEP when no IMC effects are taken 
into account (Sec. \ref{sec:MFvsVI}); the influence of the IMC in the location 
of the CEP in the presence and in the absence of the vector interaction 
(Sec. \ref{sec:IMC_CEP}). Finally we draw our conclusions in Sec. 
\ref{Conclusions}.

\section{Model and Formalism}
\label{sec:model}

The original PNJL Lagrangian \cite{PNJL} is modified in order to take into 
account the presence of an external magnetic field and the vector interaction in
(2+1) flavors:
\begin{eqnarray}
{\cal L} &=& {\bar{q}} \left[i\gamma_\mu D^{\mu}-
	{\hat m}_c \right ] q ~+~ {\cal L}_\text{sym}~+~{\cal L}_\text{det}~
  +~{\cal L}_\text{vec} \nonumber\\
&+& \mathcal{U}\left(\Phi,\bar\Phi;T\right) - \frac{1}{4}F_{\mu \nu}F^{\mu \nu},
	\label{Pnjl}
\end{eqnarray}
where the quark sector is described by the  SU(3) version of the NJL model which
includes scalar-pseudoscalar and the 't Hooft six fermion interactions that
models the axial $U_A(1)$ symmetry breaking  \cite{Hatsuda:1994pi,Klevansky:1992qe},
with ${\cal L}_\text{sym}$ and ${\cal L}_\text{det}$  given by
\begin{eqnarray}
	{\cal L}_\text{sym}= G_s \sum_{a=0}^8 \left [({\bar q} \lambda_ a q)^2 + 
	({\bar q} i\gamma_5 \lambda_a q)^2 \right ] ,
\end{eqnarray}
\begin{eqnarray}
	{\cal L}_\text{det}=-K\left\{{\rm det} \left [{\bar q}(1+\gamma_5)q \right] + 
	{\rm det}\left [{\bar q}(1-\gamma_5)q\right] \right \},
\end{eqnarray}
and a vector interaction given by \cite{Mishustin:2000ss},
\begin{equation} 
{\cal L}_\text{vec} = - G_V \sum_{a=0}^8  
\left[({\bar q} \gamma^\mu \lambda_a q)^2 + 
 ({\bar q} \gamma^\mu \gamma_5 \lambda_a q)^2 \right]. 
\label{p1} 
\end{equation}
Here, $q = (u,d,s)^T$ represents a quark field with three flavors, 
${\hat m}_c= {\rm diag}_f (m_u,m_d,m_s)$ is the corresponding (current) mass 
matrix, $\lambda_0=\sqrt{2/3}I$  where $I$ is the unit matrix in the three flavor 
space, and $0<\lambda_a\le 8$ denote the Gell-Mann matrices.
The coupling between the (electro)magnetic field $B$ and quarks, and between the 
effective gluon field and quarks is implemented  {\it via} the covariant derivative 
$D^{\mu}=\partial^\mu - i q_f A_{EM}^{\mu}-i A^\mu$ where $q_f$ represents the 
quark electric charge ($q_d = q_s = -q_u/2 = -e/3$),  $A^{EM}_\mu$ and 
$F_{\mu \nu }=\partial_{\mu }A^{EM}_{\nu }-\partial _{\nu }A^{EM}_{\mu }$ 
are used to account for the external magnetic field, and 
$A^\mu(x) = g_{strong} {\cal A}^\mu_a(x)\frac{\lambda_a}{2}$ where
${\cal A}^\mu_a$ is the SU$_c(3)$ gauge field.
We consider a  static and constant magnetic field in the $z$ direction, 
$A^{EM}_\mu=\delta_{\mu 2} x_1 B$.
In the Polyakov gauge and at finite temperature the spatial components of the 
gluon field are neglected: 
$A^\mu = \delta^{\mu}_{0}A^0 = - i \delta^{\mu}_{4}A^4$. 
The trace of the Polyakov line defined by
$ \Phi = \frac 1 {N_c} {\langle\langle \mathcal{P}\exp i\int_{0}^{\beta}d\tau\,
A_4\left(\vec{x},\tau\right)\ \rangle\rangle}_\beta$
is the Polyakov loop which is the order parameter of the $\Z_3$ 
symmetric/broken phase transition in pure gauge.

The pure gauge sector is described by an effective potential 
$\mathcal{U}\left(\Phi,\bar\Phi;T\right)$ chosen in order to reproduce the 
results obtained in lattice calculations 
\cite{Roessner:2006xn},
\begin{eqnarray}
	& &\frac{\mathcal{U}\left(\Phi,\bar\Phi;T\right)}{T^4}
	= -\frac{a\left(T\right)}{2}\bar\Phi \Phi \nonumber\\
	& &
	+\, b(T)\mbox{ln}\left[1-6\bar\Phi \Phi+4(\bar\Phi^3+ \Phi^3)
	-3(\bar\Phi \Phi)^2\right],
	\label{Ueff}
\end{eqnarray}
where 
$a\left(T\right)=a_0+a_1\left(\frac{T_0}{T}\right)+a_2\left(\frac{T_0}{T}\right)^2$, 
$b(T)=b_3\left(\frac{T_0}{T}\right)^3$.
The standard choice of the parameters for the effective potential $\mathcal{U}$ 
is $a_0 = 3.51$, $a_1 = -2.47$, $a_2 = 15.2$, and $b_3 = -1.75$.   
The parameter $T_0$ is the critical temperature for the deconfinement phase 
transition within a pure gauge approach: it was fixed to a constant $T_0=270$ MeV,
according to lattice findings. 

As a regularization scheme, we use a sharp cutoff, $\Lambda$, in three-momentum 
space, only for the divergent ultraviolet sea quark integrals.  
For the parameters of the model we consider $\Lambda = 602.3$ MeV, 
$m_u= m_d=5.5$, MeV, $m_s=140.7$ MeV, $G_s \Lambda^2= 1.835 $ and 
$K \Lambda^5=12.36$ as in \cite{Rehberg:1995kh}.
The thermodynamical potential for the three-flavor quark sector $\Omega$ is 
written as
\begin{align}
\Omega(T,\mu)=&\, G_s\sum_{f=u,d,s}\ev{\bar{q}_fq_f}^2
+4K\ev{\bar{q}_uq_u}\ev{\bar{q}_dq_d}\ev{\bar{q}_sq_s} \nonumber \\
-&\, G_V\sum_{f=u,d,s}\ev{q^{\dagger}_fq_f}^2+{\cal U}(\Phi,\bar{\Phi},T)\nonumber \\
+&\, \sum_{f=u,d,s}\pc{\Omega_{\text{vac}}^f+
\Omega_{\text{med}}^f +\Omega_{\text{mag}}^f}
\label{eq:Omega}
\end{align}
where the vacuum $\Omega^{\text{vac}}_f$, the medium $\Omega^{\text{med}}_f$, and 
the magnetic contributions  $\Omega^{\text{mag}}_f$, together with the quark 
condensates $\left\langle \bar{q_f}q_f \right\rangle$  have been evaluated with 
great detail in \cite{Menezes:2008qt.Menezes:2009uc}.
By minimizing the thermodynamical potential, Eq. (\ref{eq:Omega}), with respect 
to the order parameters $\left\langle \bar{q_f}q_f \right\rangle$, $\Phi$, and 
$\bar{\Phi}$, we obtain the mean field equations.

\begin{figure*}[tb]
	\centering
	\includegraphics[width=0.9\linewidth,angle=0]{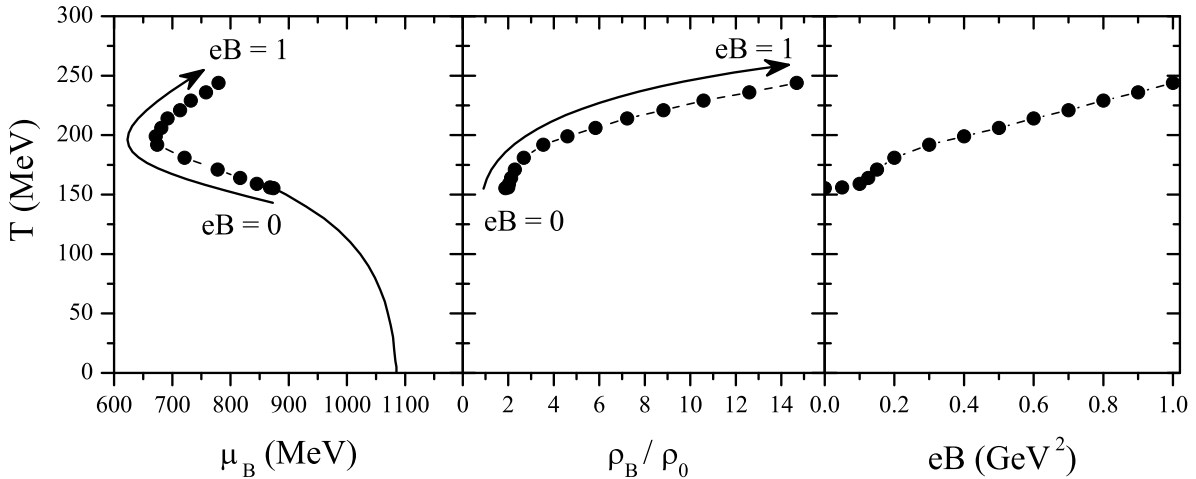}
	\caption{Location of the CEP on a diagram $T$ vs the baryonic
          chemical potential (left), vs the baryonic density (middle), and vs
          magnetic field (right) for Case IA.
          } 
	\label{fig:CEP1}
\end{figure*}

In the present study we consider the PNJL model with equal quark chemical 
potentials, $\mu_u=\mu_d=\mu_s$, which corresponds to zero charge 
(or isospin) chemical potential and zero strangeness chemical potential 
($\mu_Q = \mu_S = 0$).

In \cite{Miransky:2002rp}, it was shown that the running coupling decreases with 
the magnetic field strength. Consequently, in the NJL model the coupling $G_s$, 
which can be seen as $\propto\alpha_s$, must decrease with an increasing magnetic 
field strength. Since there is no LQCD data available for $\alpha_s(eB)$, by 
using the NJL model it is possible to fit $G_s(eB)$ in order to reproduce the 
pseudocritical chiral transition temperatures, $T_c^\chi(eB)$, obtained in LQCD 
calculations at $\mu_B=0$ \cite{baliJHEP2012}. 
The resulting fit function that reproduces the $T_c^\chi(eB)$ is 
\cite{Ferreira:2014kpa} 
\begin{equation}
  G_s(\zeta)=G_s^0\pc{\frac{1+a\,\zeta^2+b\,\zeta^3}
  {1+c\,\zeta^2+d\,\zeta^4}},
\label{eq:fit}
\end{equation}
with $a = 0.0108805$, $b=-1.0133\times10^{-4}$, $c= 0.02228$, and 
$d=1.84558\times10^{-4}$, where $\zeta=eB/\Lambda_\text{QCD}^2$ and 
$\Lambda_\text{QCD}=300$ MeV.

By using $G_s(eB)$ given by Eq. (\ref{eq:fit}) in the PNJL model both chiral 
and deconfinement transition temperatures decrease with increasing magnetic 
field strength due to the existing coupling between the Polyakov loop field and 
quarks within the PNJL model. Consequently, the coupling $G_s(eB)$ affects not only 
the chiral transition but also the deconfinement transition 
\cite{Ferreira:2014kpa}.

Next, we will study the following scenarios for the effect of a 
static external magnetic field on the location of the CEP:
\begin{enumerate}
	\item Case I with no IMC effects and the usual $G_s=G_s^0$:
		\begin{itemize}
			\item Case IA, where we take $G_V=0$ 
				(Sec.  \ref{subsec:MFvsVIa});
			\item Case IB, where we take $G_V \ne 0$, with $G_V=\alpha G_s^0$ 
				(Sec.  \ref{subsec:MFvsVIb}).
		\end{itemize}
	\item  Case II with  IMC effects, described by $G_s(eB)$ given by 
		Eq. (\ref{eq:fit}):
		\begin{itemize}
			\item Case IIA, where we take $G_V=0$, 
				(Sec. \ref{subsec:IMC_CEPa});
			\item Case IIB, with $G_V=\alpha G_s(eB)$, meaning that the stronger 
				the magnetic field the weaker the vector and scalar
				interactions, (Sec. \ref{subsec:IMC_CEPb});
			\item Case IIC, with a fixed $G_V =\alpha G^0_s$ 
			(Sec. \ref{subsec:IMC_CEPb}).
	\end{itemize}
\end{enumerate}

\section{The influences of an external magnetic field and of the vector
interaction on the location of the CEP}
\label{sec:MFvsVI}

The influence of the repulsive vector coupling on magnetized quark matter was 
investigated in \cite{Denke:2013gha} by using the SU(2) version of the NJL model.
Here, we will use the (2+1) PNJL model.
We start our study by setting $G_V=0$. In a second step we consider $G_V\ne 0$.

\subsection{ $G_V=0$}
\label{subsec:MFvsVIa}

\begin{figure*}[tb]
	\centering
	\includegraphics[width=0.49\linewidth,angle=0]{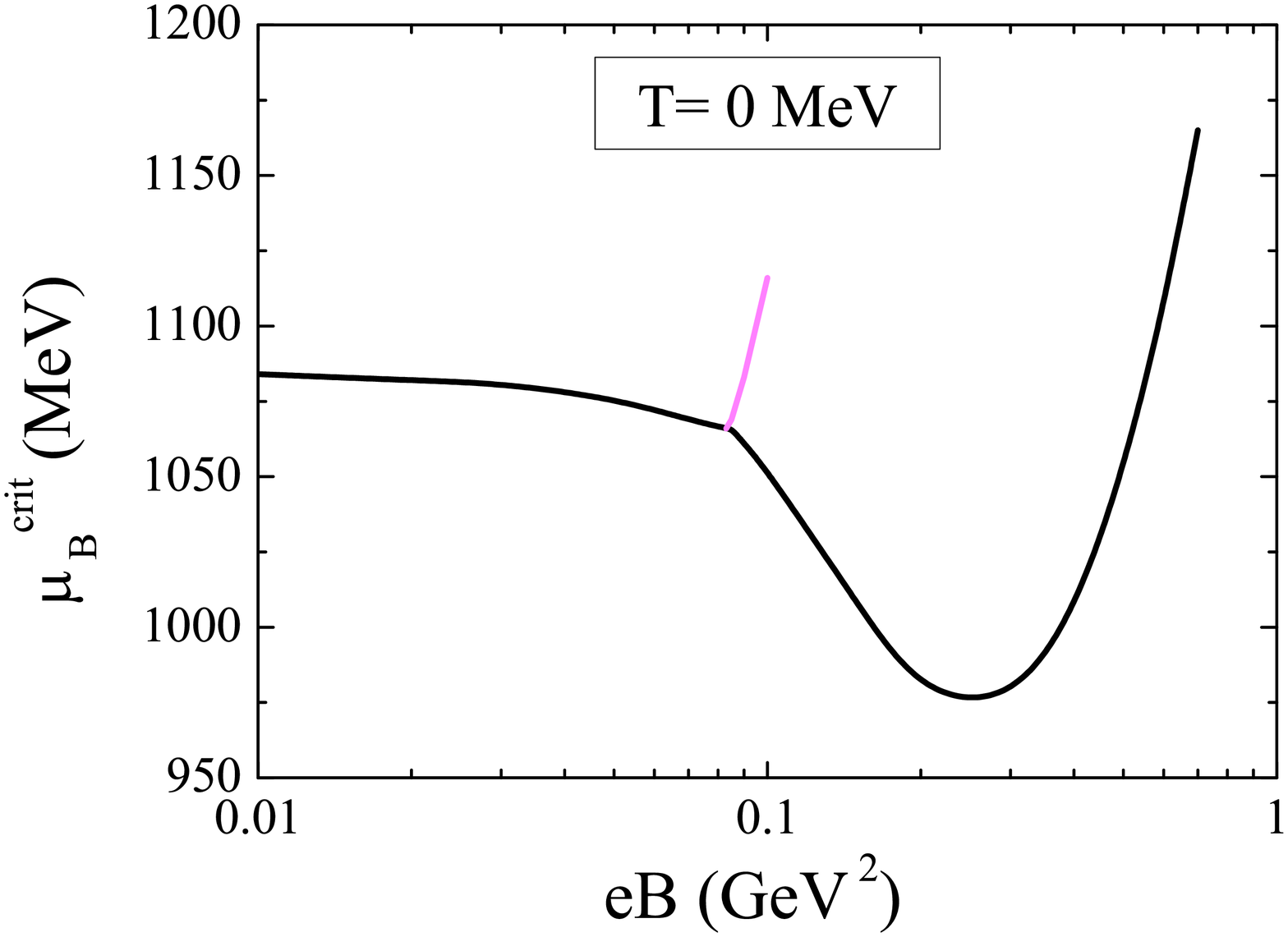}
	\includegraphics[width=0.49\linewidth,angle=0]{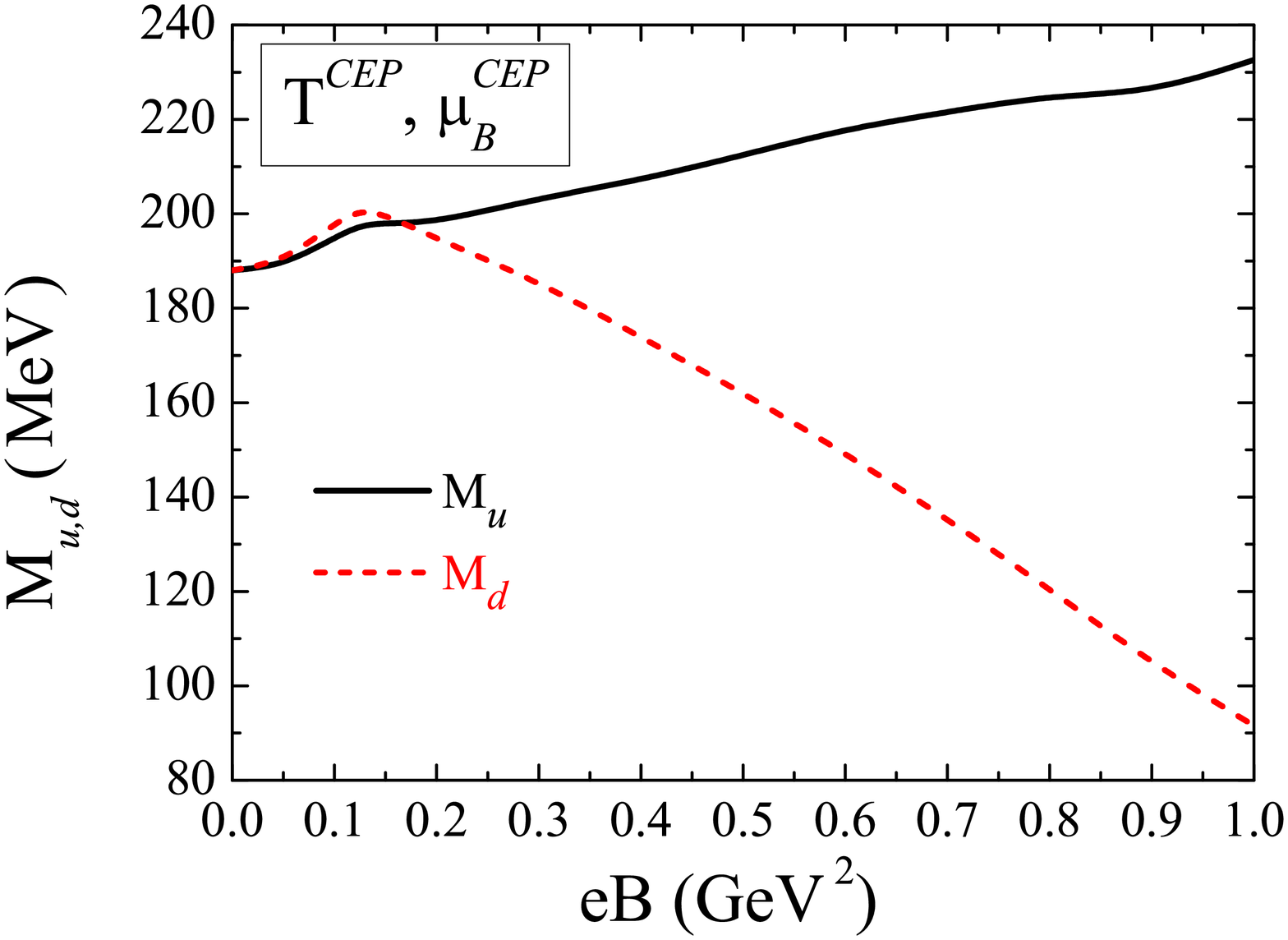}
	\caption{The critical chemical potential at $T=0$ MeV versus
          the magnetic field (left panel) and the $u$, $d$-quark
          constituent masses at the CEP as a function of the magnetic
          field intensity.
          } 
	\label{fig:CEP11}
\end{figure*}

\begin{figure*}[t]
	\centering
	\includegraphics[width=0.49\linewidth,angle=0]{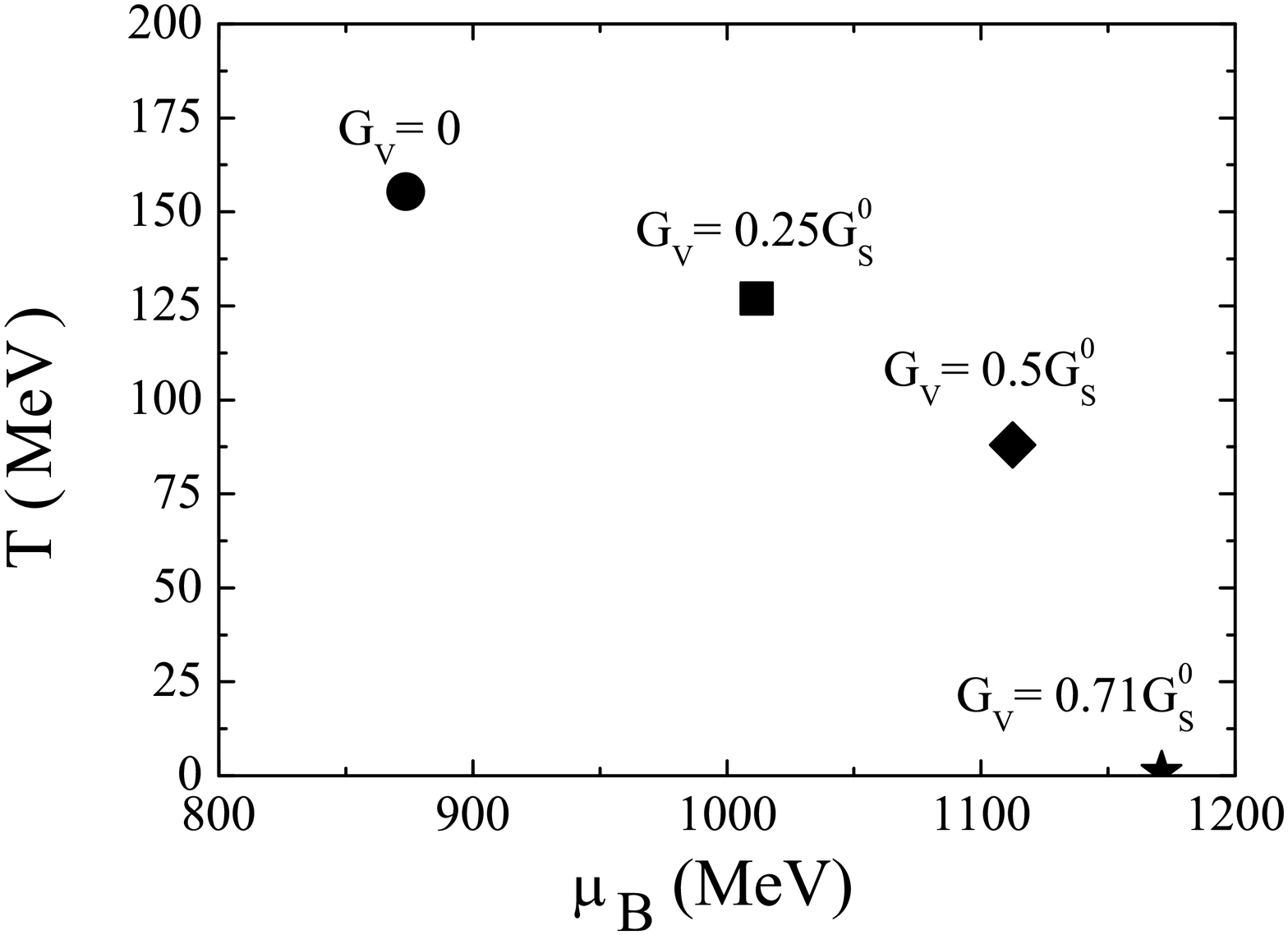}
	\includegraphics[width=0.49\linewidth,angle=0]{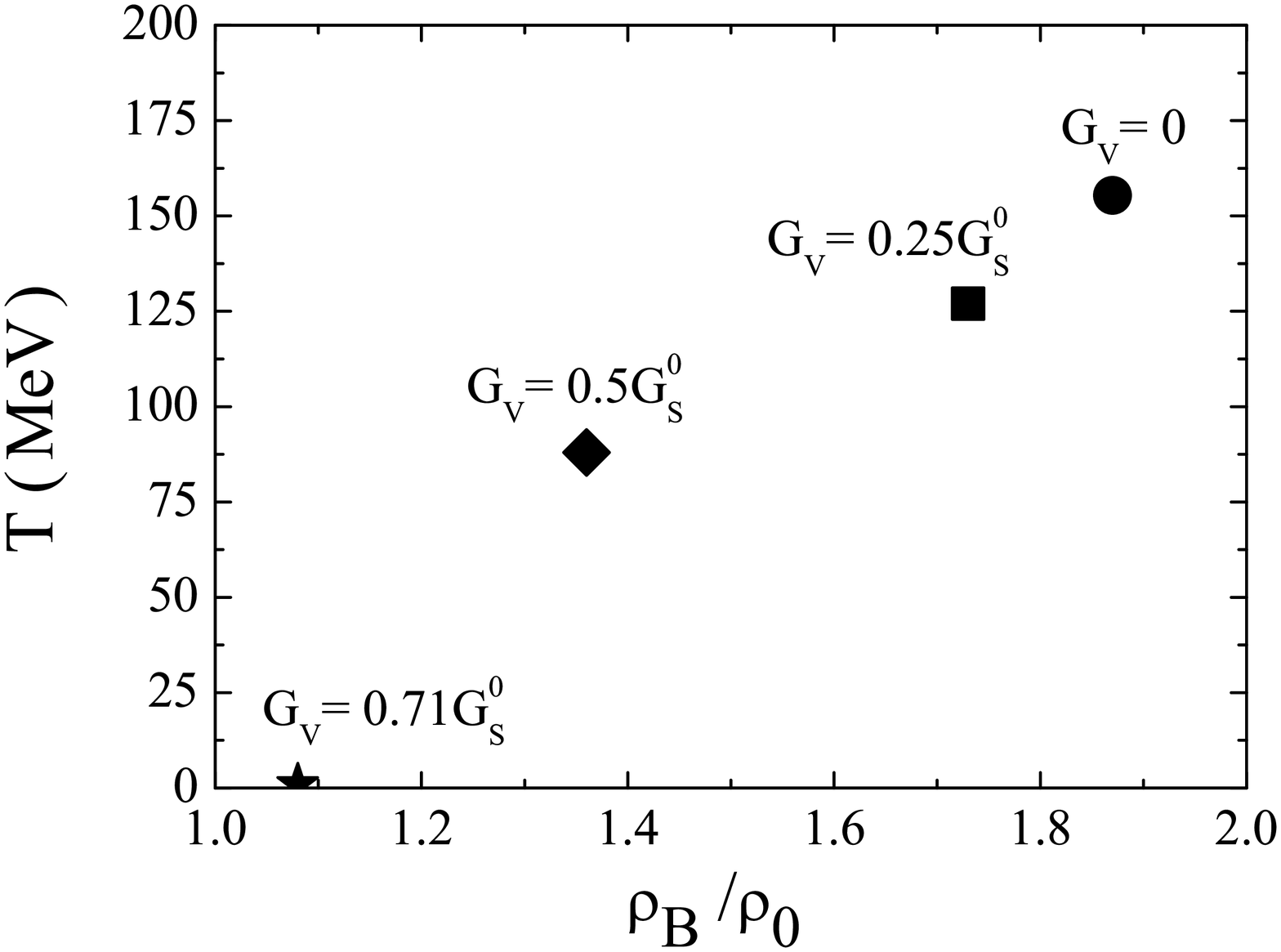}
	\caption{Effect of the strength of the vector interaction on
          the location of the CEP on a diagram $T$ vs the baryonic
          chemical potential (left) and $T$ vs the baryonic density (right).
          } 
	\label{fig:CEP2}
\end{figure*}

The $T-\mu_B$ phase diagram obtained with $G_V=0$ (Case IA) is presented in the 
left panel of Fig. \ref{fig:CEP1} and shows a trend very similar to that of the 
results previously obtained for the NJL in \cite{Avancini:2012ee}:
as the intensity of the magnetic field  increases, the temperature at which the 
CEP occurs ($T^\text{CEP}$) increases monotonically (see Fig. \ref{fig:CEP1} right 
panel) and the corresponding baryonic chemical potential ($\mu_B^\text{CEP}$) 
decreases until the critical value $eB\sim 0.4$ GeV$^2$ is reached;
for stronger magnetic fields both $T^\text{CEP}$ and $\mu_B^\text{CEP}$ increase. 
In the middle panel of Fig. \ref{fig:CEP1} the CEP is given in a $T$ versus 
baryonic density, $\rho_B/\rho_0$,  plot, and it can be seen that as the magnetic 
field increases from 0 to 1 GeV$^2$, $\rho^\text{CEP}_B$ always increases.

To understand these behaviors at finite density we start by considering
the case at $T=0$ where a first order phase transition takes place. 
In the left panel of Fig. \ref{fig:CEP11}, we present the critical chemical 
potential, $\mu_B^\text{crit}$, at which the first order phase transition occurs.
The pattern followed by $\mu_B^\text{crit}$ at $T=0$ in the PNJL model is similar, 
although for smaller values, to the one reported in \cite{Avancini:2012ee} 
at $T=1$ MeV and also at higher temperatures: slow decrease for 
$0<eB<0.06$ GeV$^2$, faster decrease until $0.12-0.18$ GeV$^2$, and monotonical 
increase afterwards.
We verify a lowering of $\mu_B^\text{crit}$ with $B$ until $eB=0.25$ GeV$^2$.
The slow decrease in $\mu_B^\text{crit}$ for increasing magnetic field strength in 
the range $0\leq eB\lesssim0.08$ GeV$^2$ is followed by a faster decrease for 
$0.08\lesssim eB\lesssim0.25$ GeV$^2$. Stronger field strengths result in a 
monotonically increasing $\mu_B^\text{crit}$. This change in behavior corresponds to 
the point where just one Landau level (LL) is filled for each flavor in the 
partially chiral restored phase. Indeed, the stronger the magnetic field, 
the larger the spacing between the levels.

At $T=\mu_B=0$ a stronger magnetic field results in an increase of the mass of 
the quarks (the increase is larger for $M_u$ than $M_d$ due to the difference in
electric charges). At finite density, however, $\mu_B^\text{crit}$ starts to decrease 
with increasing magnetic fields, indicating an easier transition to the 
partially chiral restored phase \cite{Preis:2010cq}. 
This result was already seen in \cite {Avancini:2012ee}. 
For $eB$ above $0.25$ GeV$^2$, $\mu_B^\text{crit}$ increases.  

Also noteworthy to point out is the existence of a range of magnetic fields, 
$0.083\lesssim eB\lesssim0.1$ GeV$^2$, where at least two first order phase 
transitions occur (see Fig. \ref{fig:CEP11}, left panel\footnote{Around 
$eB\approx 0.085$ GeV$^2$ a small third phase transition 
(not visible on Fig. \ref{fig:CEP11}) can be found on a very small range.}), 
in accordance with what was found in the SU(2)  
\cite{Denke:2013gha,Allen:2013lda} and SU(3) NJL models \cite{Grunfeld:2014qfa}. 
This cascade of transitions will result in the existence of multiple CEPs at 
finite temperature.
The CEP on which we focus most of our attention in the present and next sections 
is the one that subsists to the highest temperature.

As was discussed above, in the weak magnetic field regime an increasing magnetic 
field results in a smaller critical chemical potential for the first order 
transition at $T=0$, even if the quarks' masses have already started to increase. 
As this corresponds to a shift of the first order transition line toward a
smaller chemical potential, the observed decrease in $\mu_B^\text{CEP}$ 
follows naturally. This effect is dominant over that of the increase of the 
quark masses at the CEP (both quark masses at the CEP increase with magnetic 
field strength for $eB\lesssim 0.125$ GeV$^2$) which should hinder the first 
order partial chiral restoration (see Fig. \ref{fig:CEP11}, right panel).
A similar behavior is also obtained within the NJL model used in 
\cite{Avancini:2012ee}.

Above a critical strength for the magnetic fields, $eB\gtrsim 0.125$ GeV$^2$, 
there is a clear asymmetry in the CEP quark mass response to an increasing 
magnetic field strength: a strong decrease in $M_d$ as opposed to the smooth 
increase in $M_u$ (due to the charge difference the $d$-quark coupling to the 
magnetic field is weaker). This behavior is accompanied by an increase of the 
baryonic density at which the CEP occurs (Fig. \ref{fig:CEP1}, right panel). 

For stronger magnetic fields ($eB\gtrsim 0.4$ GeV$^2$) both $T^\text{CEP}$ and 
$\mu_B^\text{CEP}$ increase.
This can be understood as a result of a decreasing number of occupied
LL due to the large intensity of the field and the greater difficulty 
in restoring chiral symmetry.

\subsection{ $G_V\ne 0$}
\label{subsec:MFvsVIb}

The role of the vector interaction in the PNJL was studied in  detail in 
\cite{Fukushima:2008wg,Friesen:2014mha}. The main conclusion was that the CEP 
can be absent when the value of the coupling $G_V$ is greater than a critical 
value $G_V^\text{crit}$. With the present parametrization this critical value 
is when $\alpha \approx 0.71$, i.e, $G_V^\text{crit}\approx 0.71G_s^0$ 
(see both panels of  Fig. \ref{fig:CEP2}). 
As the value of the coupling $G_V$ is increased from 0 to $G_V^\text{crit}$ the first 
order phase transition is weakened and the CEP occurs at lower temperatures and 
larger chemical potentials but smaller densities. 

\begin{figure*}[tb]
	\centering
	\includegraphics[width=0.49\linewidth,angle=0]{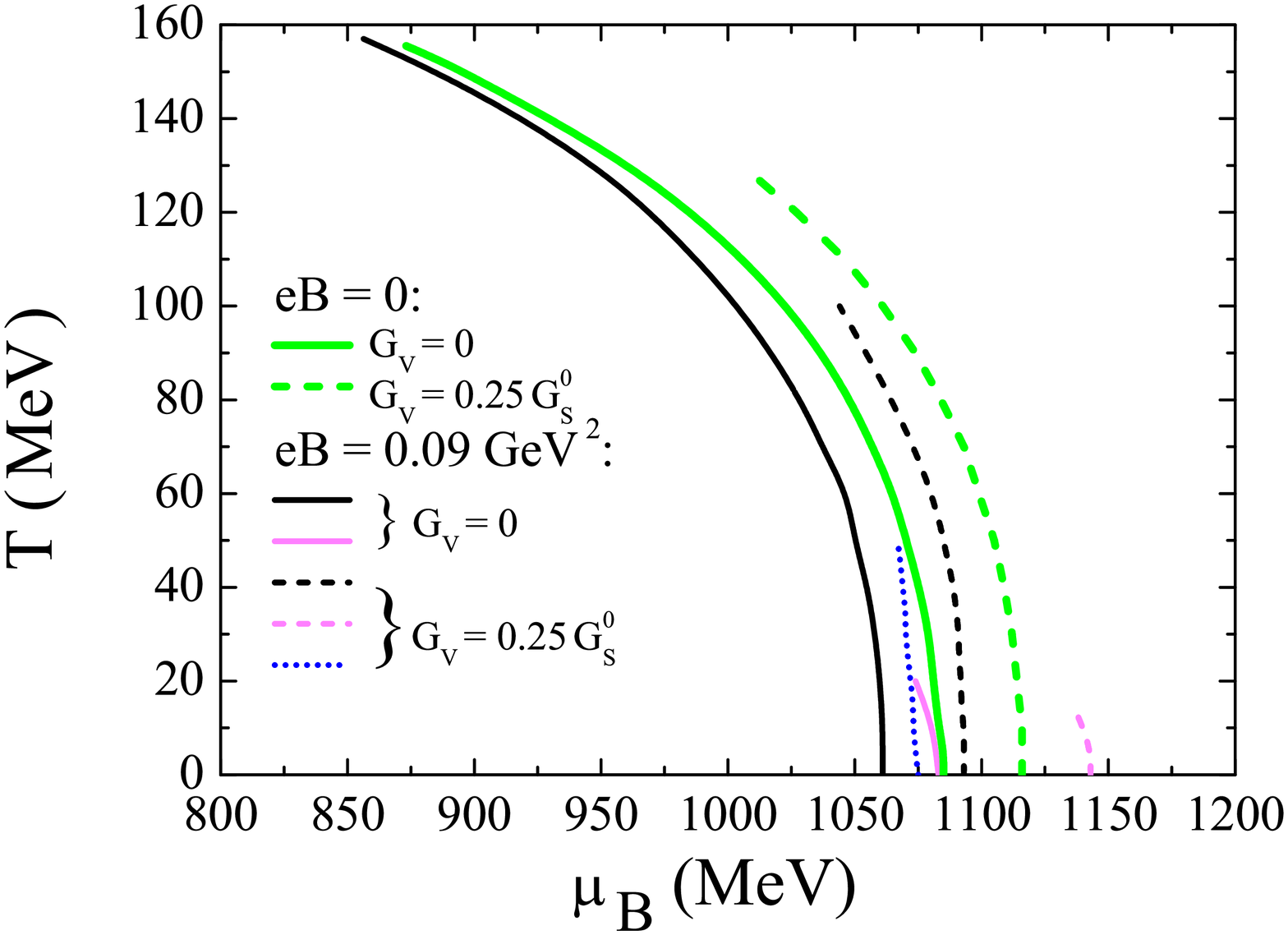} 
	\includegraphics[width=0.49\linewidth,angle=0]{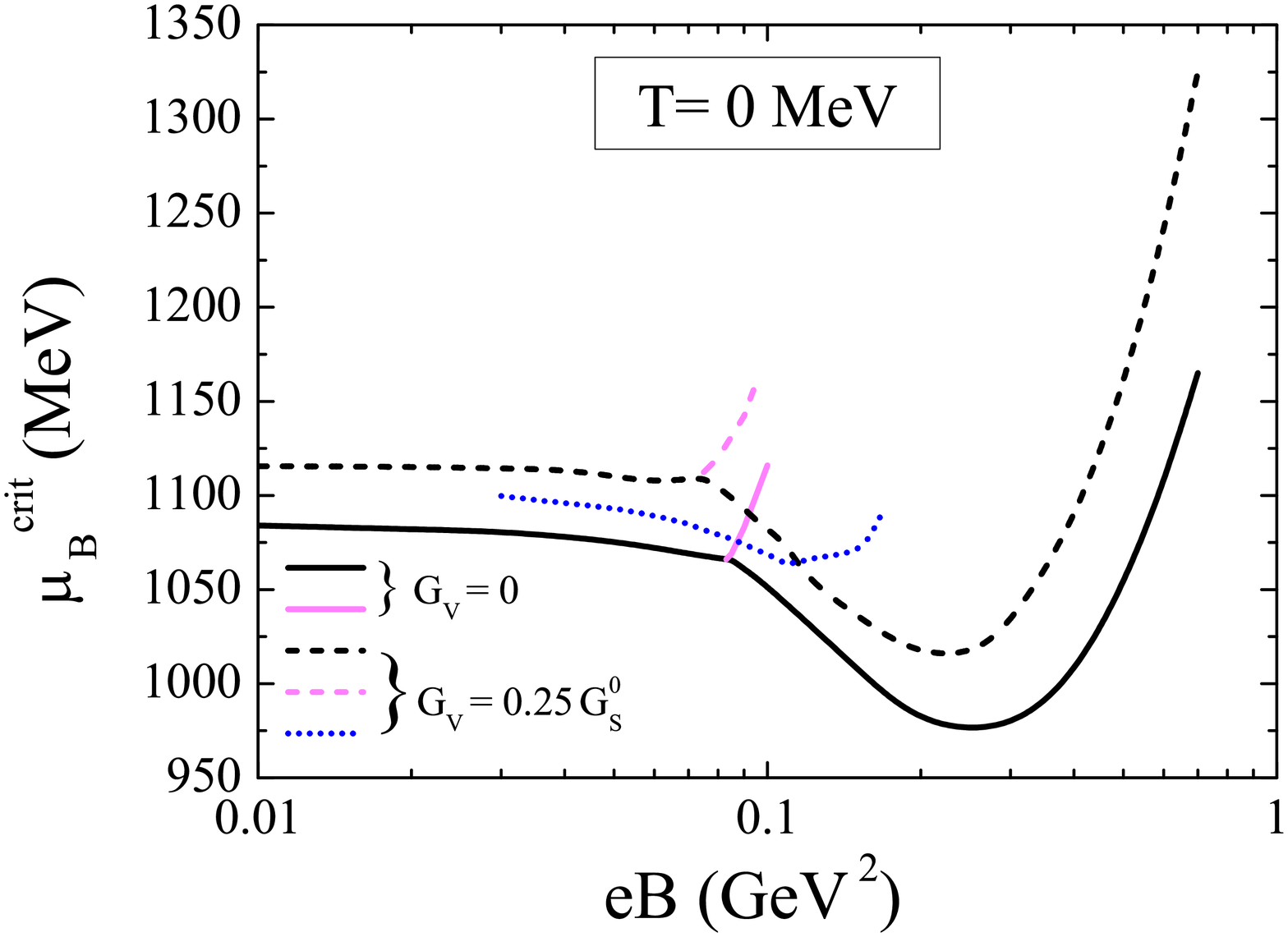}
	\caption{QCD phase diagram in a $T-\mu_B$ plot for Cases IA
          and IB, not including IMC effects,  discussed in the text, left panel. 
					The critical chemical potential as a function of the magnetic field
          intensity at $T=0$, right panel. 
					The green thick lines correspond to no magnetic field, and dashed (full) 
					lines were obtained with (without) the vector interaction.
          } 
	\label{fig:CEP_eB_GV_1}
\end{figure*}

When the external magnetic field and the vector interaction are taken into 
account simultaneously, the scenario becomes more complex. 
In our discussion we will fix $\alpha=0.25$ ($G_V=0.25G_s^0$) and $eB=0.09$ GeV$^2$ 
to compare to cases with $G_V=0$ and $G_V=0.25G_s^0$ at zero magnetic field.
The results are presented in Fig. \ref{fig:CEP_eB_GV_1}, left panel. In the
following we will show that for the light sector the (2+1) PNJL model is in 
agreement with  the SU(2) NJL model (see \cite{Denke:2013gha} for details).

We start by exploring in detail the phase diagram in the presence of a 
magnetic field  $eB=0.09$ GeV$^2$ at $G_V=0$ where two first order phase 
transitions in cascade occur at $T=0$.  We identify a decrease of $\mu_B^\text{crit}$ 
for both transitions (see black and red full lines of Fig. 
\ref{fig:CEP_eB_GV_1} left panel) when compared with the transition at zero 
magnetic field (green full line in the same figure).

At $T=0$ these transitions occur for a given $\mu_B$ at which the gap equations 
have two stable solutions ($M_i^{I}$ and $M_i^{II}$) leading to the coexistence 
of two different quark masses at the same pressure and temperature.
The first transition occurs at $\mu_B = 1061$ MeV, being 
$M_{u(d)}^{I- 1\text{st}}=378.7\,(374.8)$ MeV and $M_{u(d)}^{II- 1\text{st}}=166.0\,(158.9)$ 
MeV, and the second transition occurs at $\mu_B = 1083$ MeV, where 
$M_{u(d)}^{I- 2\text{nd}}=149.8\,(141.4)$ MeV and $M_{u(d)}^{II- 2\text{nd}}=49.4\,(49.0)$ MeV.

At this point some other relevant aspects of these results are worth highlighting:
\begin{enumerate}[label=\roman*)] 
\item At $\mu_B=0$ and $eB=0.09$ GeV$^2$ we have the magnetic catalysis effect 
as $M_{u}$ and $M_{d}$ are higher than the respective vacuum values 
($M_{u}^\text{vac}=M_{d}^\text{vac}=367.7$ MeV); 
\item At the first phase transition the values of 
$M_{u(d)}^{II- 1\text{st}}=166.0\, (158.9)$ MeV are still far from the respective 
current values ($m_u=m_d=5.5$ MeV), and a second transition is needed to bring 
$M_{u(d)}^{II- 2\text{nd}}$ to values closer to $m_{u(d)}$ which is consistent with a 
region where the chiral symmetry is partially restored.
\item At the second phase transition $M_{u(d)}^{II- 2\text{nd}}=49.4\,(49.0)$ MeV and 
these values are already smaller than $M_{u}^{II}=M_d^{II}=52.2$ MeV at $eB=0$.
\item If we take the chiral limit for the light sector, $m_u=m_d=0$, the 
restoration of chiral symmetry, $M_u=M_d=0$, does not coincide with the first 
phase transition:
at $\mu_B=1012$ MeV there is a jump in the quarks' masses from 
$M_{u(d)}^{I- 1\text{st}}=366.5\,(362.5)$ MeV to $M_{u(d)}^{II- 1\text{st}}=36.1\, (33.5)$ MeV 
but only at $\mu_B=1017$ MeV we have $M_u=M_d=0$.
This is a direct manifestation of the condensate enhancement by magnetic catalysis.
\item At finite temperature the second transition exists only at low temperatures 
and turns into a CEP at around $T\sim20$ MeV (see red full line, left panel 
of Fig. \ref{fig:CEP_eB_GV_1}). Increasing the temperature, the remaining 
first order transition will subsist until $T^\text{CEP}=157$ MeV and 
$\mu_B^\text{CEP}=856$ MeV, closer to the CEP for $eB=0$ 
($T^\text{CEP}=156$ MeV, $\mu_B^\text{CEP}=873$ MeV \cite{Costa:2013zca}).
\end{enumerate}

When the vector interaction is also taken into account in the presence of a 
magnetic field, two competing effects come into play: on the one hand, an increase 
of $G_V$ at $eB=0$ weakens the first order phase transition which leads to the 
disappearance of the CEP in the $\mu_B$ axis (this can be seen in the left panel 
of Fig. \ref{fig:CEP_eB_GV_1} where the green dashed line shows how $G_V$ affects 
the first order phase transition);
on the other hand, an increase of the magnetic field at $G_V=0$ has an 
opposite influence in the first order transition and CEP location,
moving the transition line to smaller chemical potentials, at least until
$eB=0.4$ GeV$^2$, as shown by the black line ($eB=0.09$ GeV$^2$) in the left 
panel of Fig. \ref{fig:CEP_eB_GV_1}.   
Besides, as we already saw, the presence of a strong enough magnetic field can 
drive multiple CEPs due to the existence of several first order transitions at 
$T=0$.

\begin{figure*}[tb]
\centering
		\includegraphics[width=0.49\linewidth,angle=0]{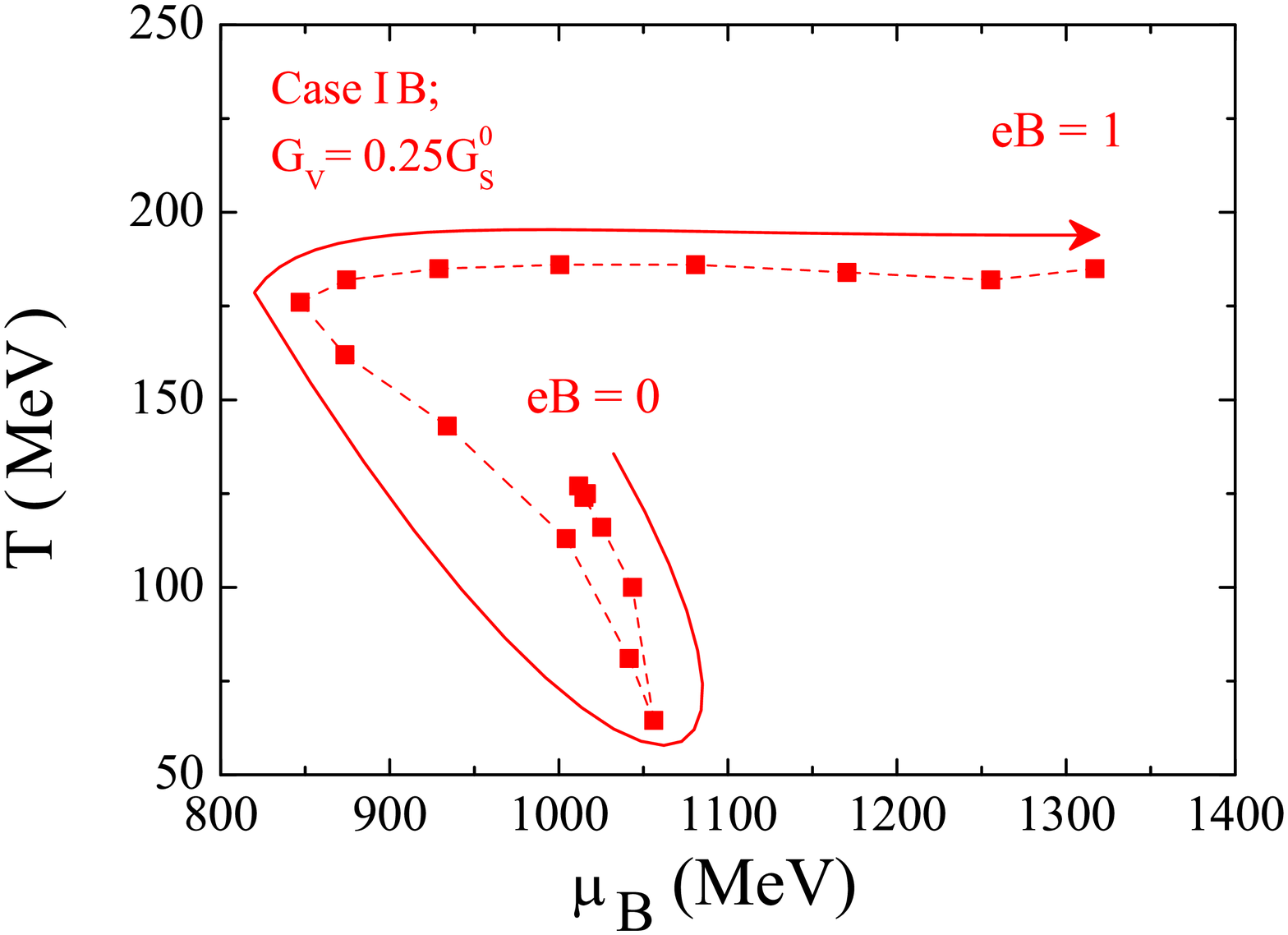}
    \includegraphics[width=0.49\linewidth,angle=0]{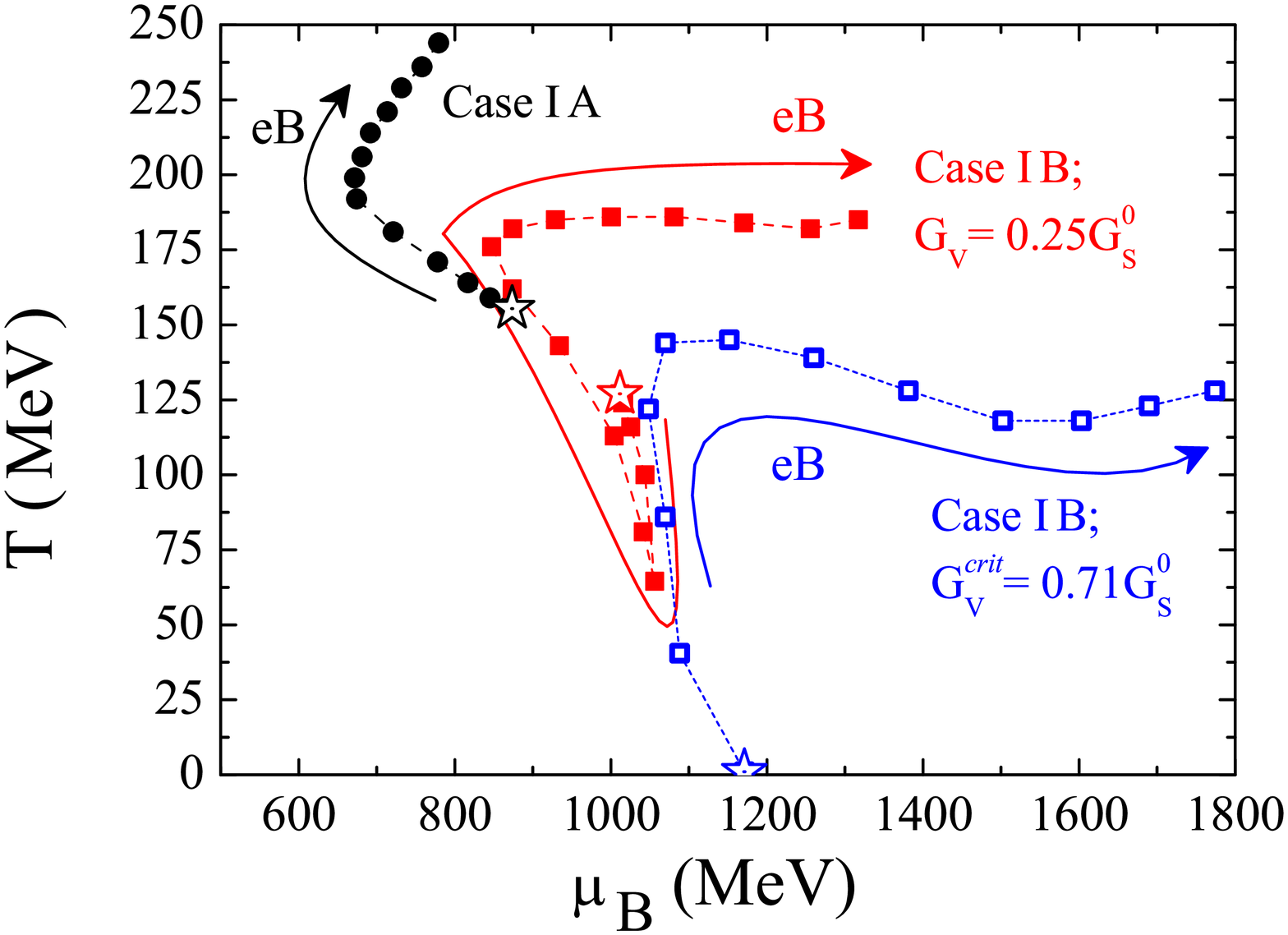}
    \caption{The location of the CEP for Case IB, with the inclusion of the 
            vector interaction and no IMC, the results for Case IB with 
            $\alpha=0.25$ and 0.71 are compared with Case IA, with no vector 
            interaction. The thin arrows indicate the direction of increasing 
            magnetic field intensity.
            }
\label{fig:CEP_eB_GV_2}
\end{figure*}

Now, we discuss the case $G_V=0.25G_s^0$ starting with $T=0$ in the presence
of an external magnetic field.
As $B$ increases, several first order transitions in 
cascade take place, similar to the $G_V=0$ results.
The  critical chemical potential  at which the transitions occur is shown in 
the right panel of Fig. \ref{fig:CEP_eB_GV_1} (dashed lines): there is a range 
of magnetic fields, $0.03\lesssim eB\lesssim0.11$ GeV$^2$, where two transitions 
occur and a range, $0.07\lesssim eB\lesssim0.09$ GeV$^2$, where three transitions 
in cascade coexist. A first conclusion about the combined effect of $B$ and 
$G_V$ is the appearance of intermediate transitions for a wider range of 
magnetic fields.

Let us fix once again $eB=0.09$ GeV$^2$, a scenario where we have three phase 
transitions:  compared with the $eB=0$ result (green dashed line),  two
transitions occur at lower $\mu_B$ 
(black and blue dashed lines in Fig. \ref{fig:CEP_eB_GV_1}, left panel)
and a third one at higher $\mu_B$ (red dashed line in the same panel). 
At finite temperatures the phase transitions will give rise to three
CEPs. The  CEP with the larger temperature,
$T^\text{CEP}=100$ MeV and $\mu_B^\text{CEP}=1044$ MeV, 
(black dashed line  in Fig. \ref{fig:CEP_eB_GV_1}, left panel), has a 
lower temperature and larger chemical potential than the CEP at $eB=0$.

In Fig. \ref{fig:CEP_eB_GV_2} left panel, the location of the CEP
obtained with $G_V=0.25\, G_s$ is plotted for different values of the
magnetic field, the thin arrowed line shows the direction of
increasing fields. Going along the direction of increasing fields,
there is a  first region corresponding to the weaker magnetic fields $B<0.1$ 
GeV$^2$, where the CEP temperature decreases and CEP
chemical potential  increases. Next, for $0.1< eB\lesssim 0.4$
GeV$^2$ the CEP temperature increases and the CEP chemical potential
decreases, and finally for stronger fields the CEP chemical potential
increases while the CEP temperature remains practically unchanged.

For $B<0.1$ GeV$^2$, $\mu_B^\text{CEP}$ increases slightly and  the
respective baryonic density decreases, the magnetic field  having an effect that 
adds to the one of $G_V$. In this  region $G_V$ is  dominant and weakens 
the first order phase transition.
This effect was clarified in \cite{Denke:2013gha}: although the effect of $G_V$ 
goes always in the direction of reducing the density at the first order phase 
transition, the magnetic field does not present a monotonic behavior as seen 
for the scenario without the vector term. 

Above $B=0.1$ GeV$^2$, the  effect of $B$ on the CEP location is close to the 
one obtained for $G_V=0$, except that for $eB\gtrsim0.4$ GeV$^2$ the CEP 
temperature keeps increasing  if $G_V=0$, while for a finite $G_V$ this trend is 
strongly reduced and the temperature does not change much; compare black and red 
curves of  Fig. \ref{fig:CEP_eB_GV_2} right panel.

\subsection{ $G_V=G_V^\text{crit}$}

Finally, we investigate the case $G_V^\text{crit}\approx 0.71G_s^0$. 
The results are presented in the right panel of Fig. \ref{fig:CEP_eB_GV_2}
(blue curves). 
The effect of the field is to restore a first order phase transition. 

For $0.01\lesssim eB\lesssim0.1$ GeV$^2$ a very complex structure of multiple 
first order transitions appears at $T=0$.
These multiple transitions will be washed out with the increasing of the 
temperature until just one CEP remains. However, more than one CEP can occur
at very close temperatures like the scenario found in \cite{Costa:2013zca}.
Above 0.1 GeV$^2$ this complex structure disappears 
(in the right panel of Fig. \ref{fig:CEP_eB_GV_2} we start to represent CEP 
for $eB >0.1$ GeV$^2$). 
For $0.1\lesssim eB\lesssim0.4$ GeV$^2$ we verify that the larger $eB$ the larger 
$T^\text{CEP}$ until a maximum $T\sim 145$ MeV is reached. In the same range, the CEP 
chemical potential decreases slightly for $eB <0.3$ GeV$^2$, and increases
for stronger fields. 

Increasing further the magnetic field, i.e.  $eB >0.4$ GeV$^2$, the
$\mu_B^\text{CEP}$ increases while the temperature does not change much until 
$eB=1$ GeV$^2$, showing also stabilization of the CEP temperature 
even if weaker than the one obtained for $G_V=0.25\, G_s$. 
The number of occupied LL becomes quite  small.
Indeed, taking $eB=0.4$ GeV$^2$, the  number of occupied LL at the CEP 
decreases with increasing $G_V$. 
The behavior obtained is the result of a clear competition between the
magnetic field that disfavors chiral symmetry restoration and the 
vector interaction with an opposite effect.

\section{The influence of the inverse magnetic catalysis in the location of the 
critical end point}
\label{sec:IMC_CEP} 

In this section we will investigate the influence, in the location of the 
CEP, of the IMC effect observed in LQCD calculations at zero chemical potential. 
First, the discussion will not include the vector interaction, Case IIA  
(Sec. \ref{subsec:IMC_CEPa}) and next the IMC effects will be considered
together with the vector interaction, Cases IIB and IIC (Sec.  
\ref{subsec:IMC_CEPb}).

\begin{figure*}[th]
\centering
	\includegraphics[width=0.9\linewidth,angle=0]{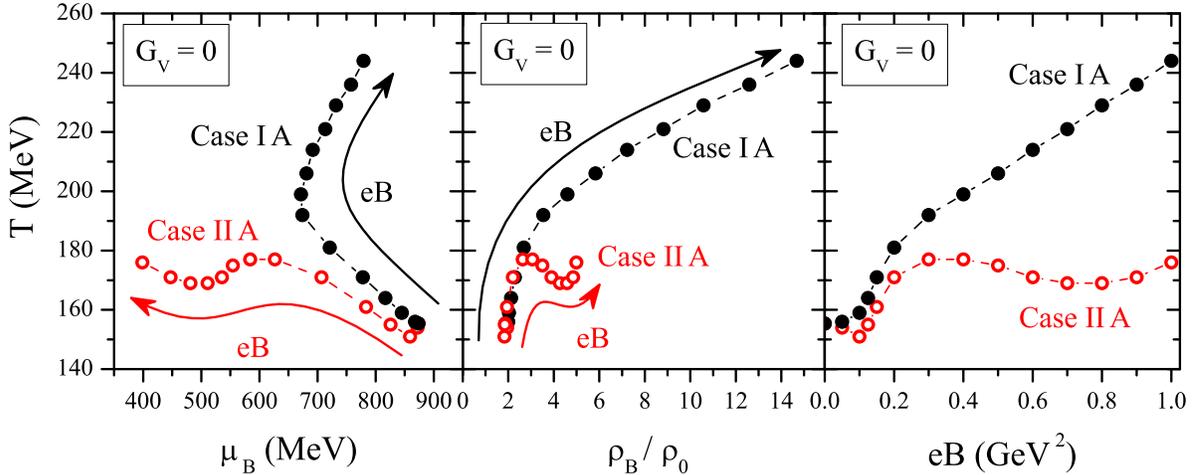}
	\caption{Location of the CEP, $T^\text{CEP}$ versus $\mu_B^\text{CEP}$
          (left panel) and $T^\text{CEP}$ versus $\rho_B^\text{CEP}$
          (right panel), for different intensities of the magnetic
          field and  excluding the vector interaction, without IMC effects 
          $G_s=G_s^0$ (red curve) and  with IMC effects $G_s=G_s(eB)$ 
          (black curve).}
\label{fig:CEP_IMC}
\end{figure*}

\subsection{$G_V=0$}
\label{subsec:IMC_CEPa} 

\begin{figure*}[tb]
	\centering
	\includegraphics[width=0.49\linewidth,angle=0]{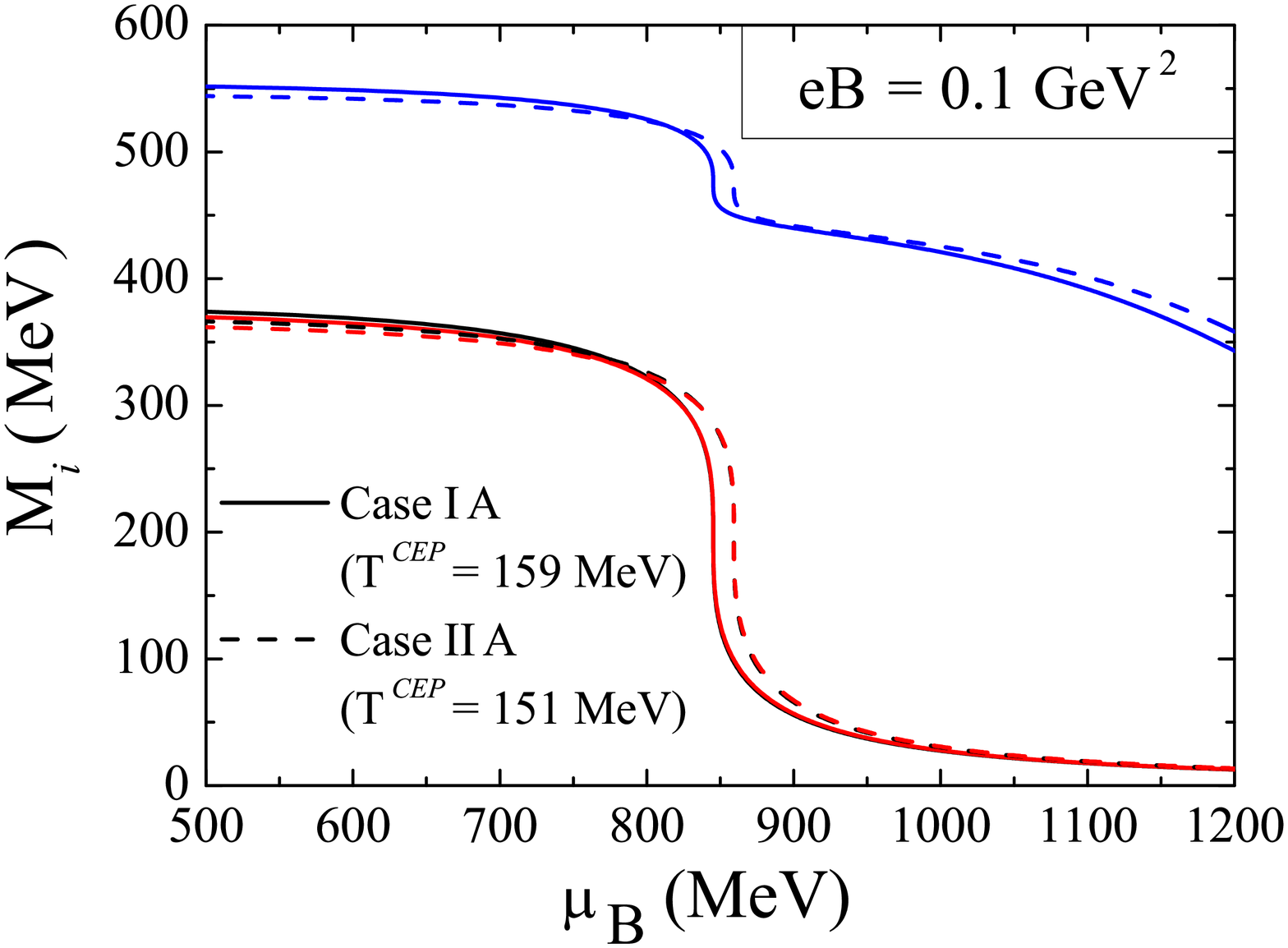} 
	\includegraphics[width=0.49\linewidth,angle=0]{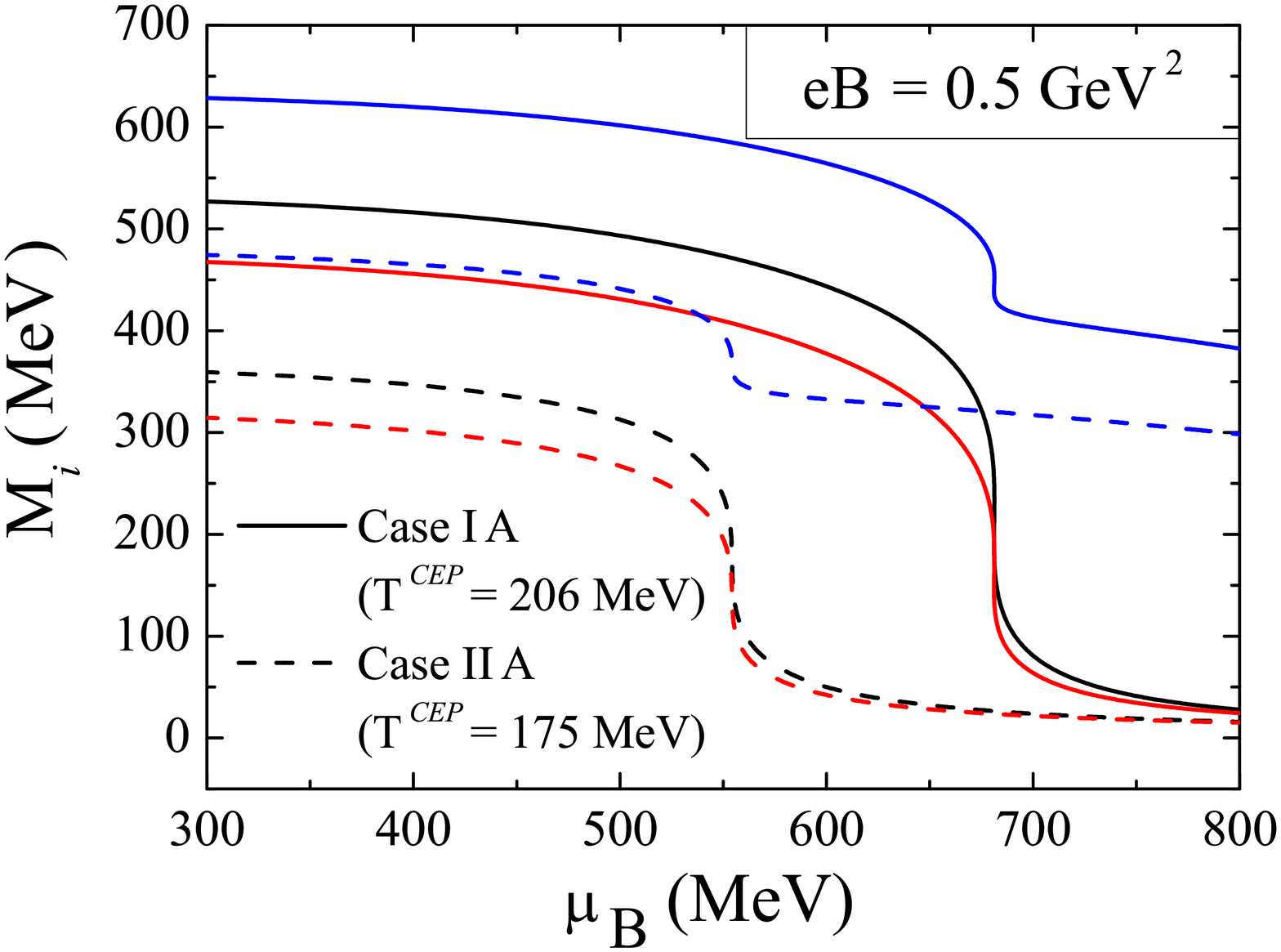} 
	\caption{Masses of the quarks as a function of $\mu_B$ for two
          intensities of the magnetic field: $eB=0.1$ GeV$^2$ (left)
          and : $eB=0.5$ GeV$^2$ (right), at the respective
          $T^\text{CEP}$.
          } 
\label{CEP_mass}
\end{figure*}

The effect of the IMC on the CEP location  excluding the vector
interaction, Case IIA,  is presented in Fig. \ref{fig:CEP_IMC} (red points) 
in the $T-\mu_B$ plane (left panel) and in the $T-\rho_B/\rho_0$ plane 
(middle panel), for different intensities of the magnetic field, and in the 
$T-eB$ plane (right panel). 
For comparison we include in the same figure the CEP location without IMC 
effects, Case IA (black curve).
We clearly observe a different behavior between these two scenarios: 
at $B=0$ both CEPs coincide but, already for small values of $B$, the IIA CEP is 
moved to lower temperatures and chemical potentials, keeping, however,
a similar behavior to IA until $eB\sim 0.3$~GeV$^2$. 
The large differences start for stronger magnetic fields: 
in Case IIA the position of the CEP oscillates between $T\approx 169$ and 
$T\approx 177$ MeV while the chemical potential takes increasingly smaller 
values; in Case IA both values of $T$ and $\mu_B$ for the CEP increase 
(see black curve, left panel of Fig. \ref{fig:CEP_IMC}).
In the middle panel of Fig. \ref{fig:CEP_IMC} the position of  the CEP in the 
$T-\rho_B/\rho_0$ plane is presented. Comparing  Cases IA  and  IIA, it
is found that the IMC effect on the  CEP results on its shift  to smaller 
temperatures and densities especially for higher values of the magnetic field.

The reason for these behaviors lies in the fact that the weakening of the 
coupling $G_s(eB)$ will make the restoration of chiral symmetry easier. 
Increasing the magnetic field is not sufficient to counteract this effect, as can 
be seen in Fig. \ref{CEP_mass} where we plot the quark masses ($M_u$: black line; 
$M_d$: red line; $M_s$: blue line) as a function of $\mu_B$ for the respective 
$T^\text{CEP}$ at $eB=0.1$ and $eB=0.5$ GeV$^2$. 
At $eB=0.1$ GeV$^2$, left panel, $G_s$ is barely affected by the magnetic field 
when IMC effects are included and the values of the quark masses are  very close 
to each other for both cases: in Case IIA the CEP occurs at smaller temperatures 
and at near, slightly higher, chemical potentials.
When $eB=0.5$ GeV$^2$, right panel, the quark masses in Case IA have increased 
with respect to the $eB=0$ case (due to the MC effect), making the restoration of 
chiral symmetry more difficult to achieve.
However, when $G_s=G_s(eB)$, Case IIA, the masses of the quarks are smaller 
than their $eB=0$ value (due to IMC effect), leading to a faster restoration of 
chiral symmetry at small temperatures and chemical potentials. 

Eventually, with the increase of $eB$ the CEP would move toward $\mu_B=0$, and 
the deconfinement and chiral phase transitions would always be of first order.

\begin{figure*}[t!]
\centering
    \includegraphics[width=0.45\linewidth,angle=0]{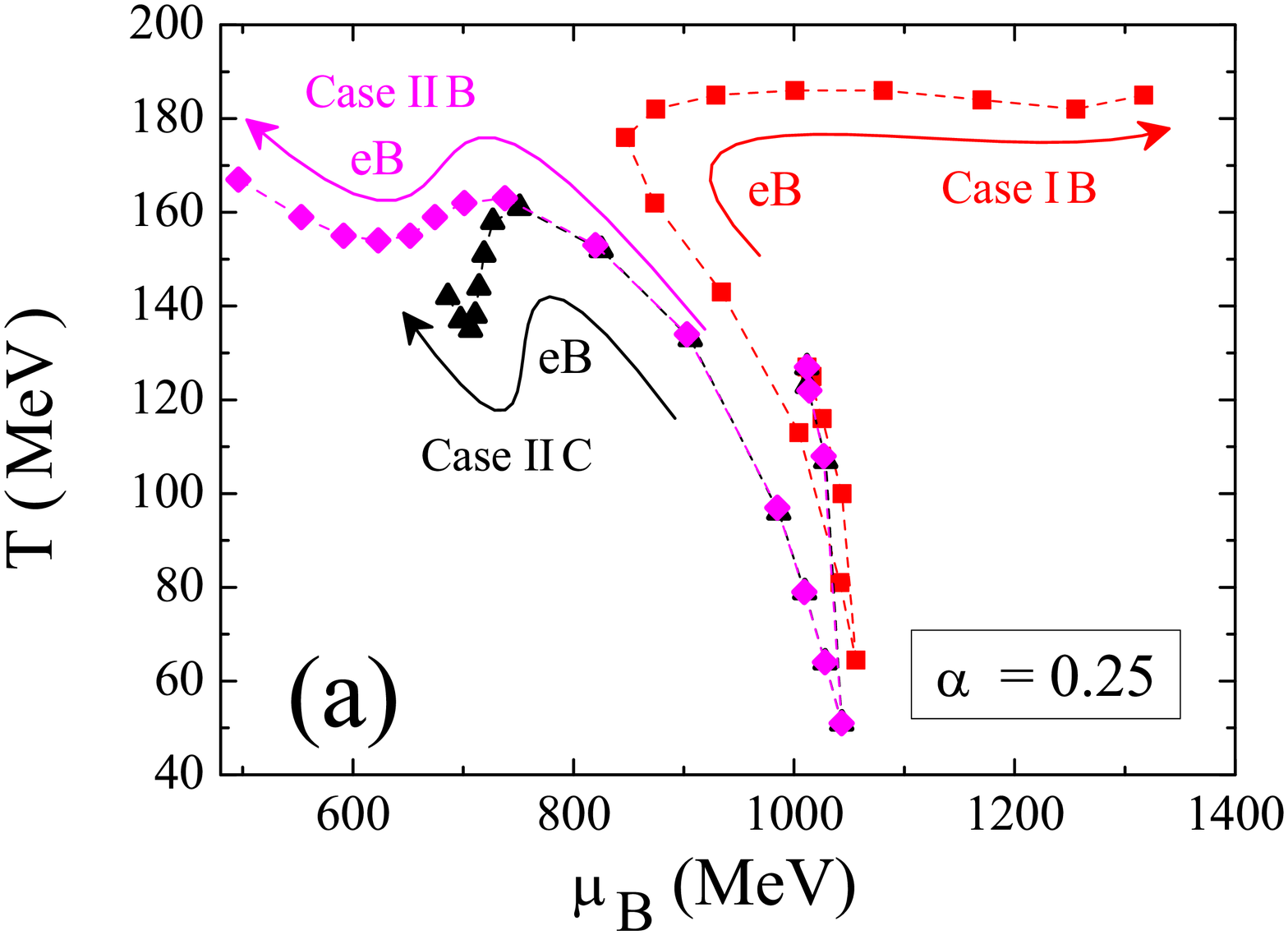}
    \includegraphics[width=0.45\linewidth,angle=0]{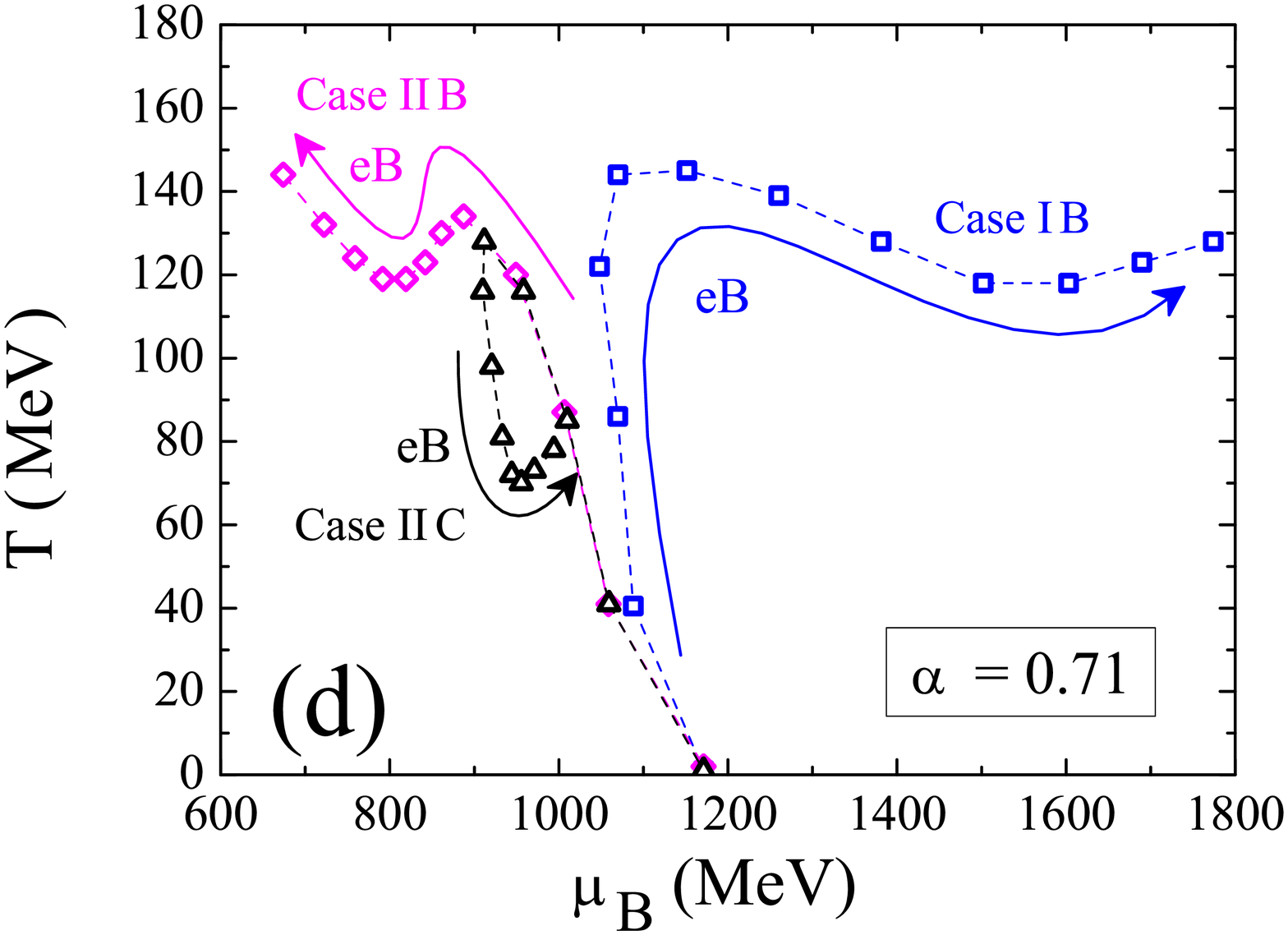}
    \includegraphics[width=0.44\linewidth,angle=0]{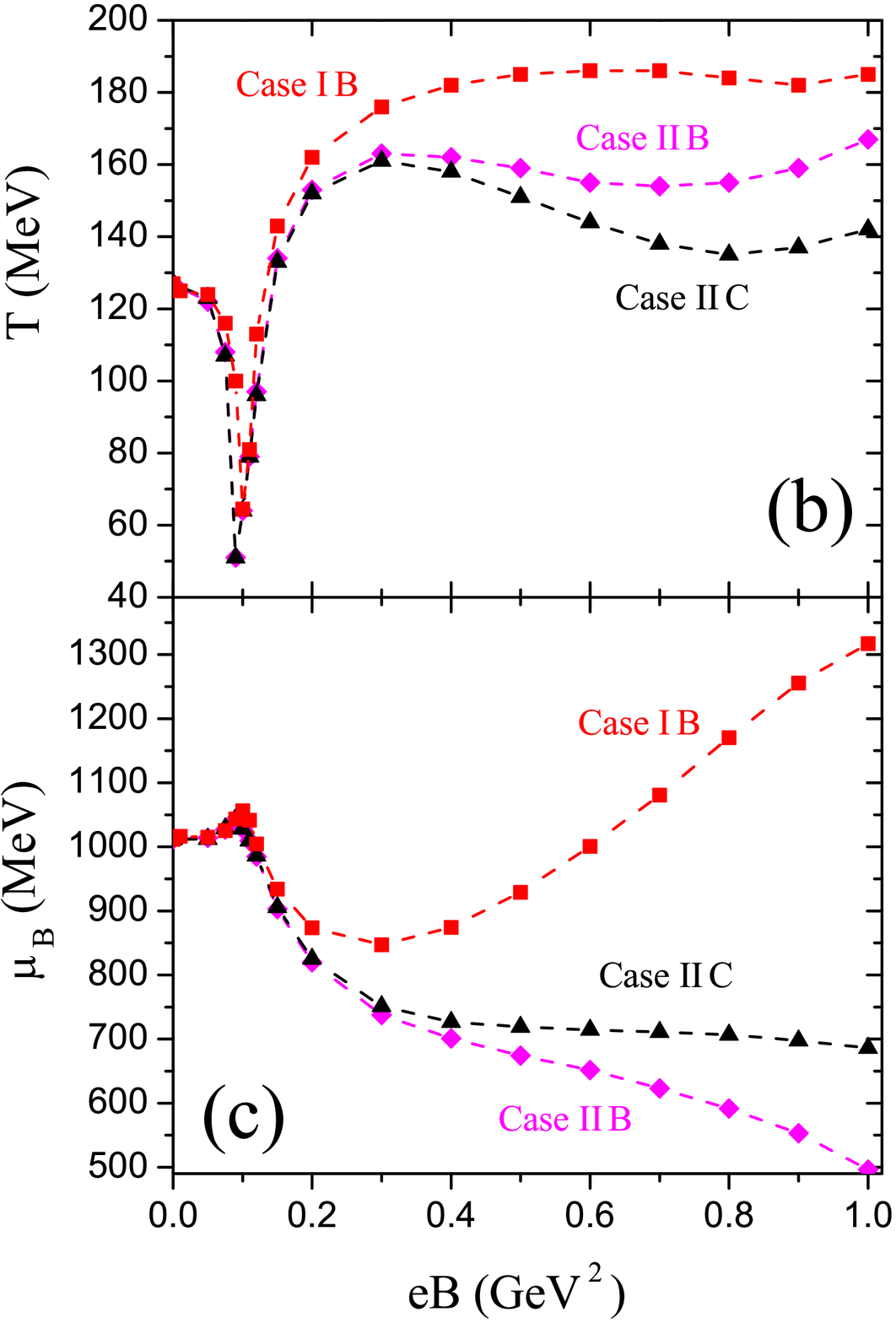}
    \includegraphics[width=0.44\linewidth,angle=0]{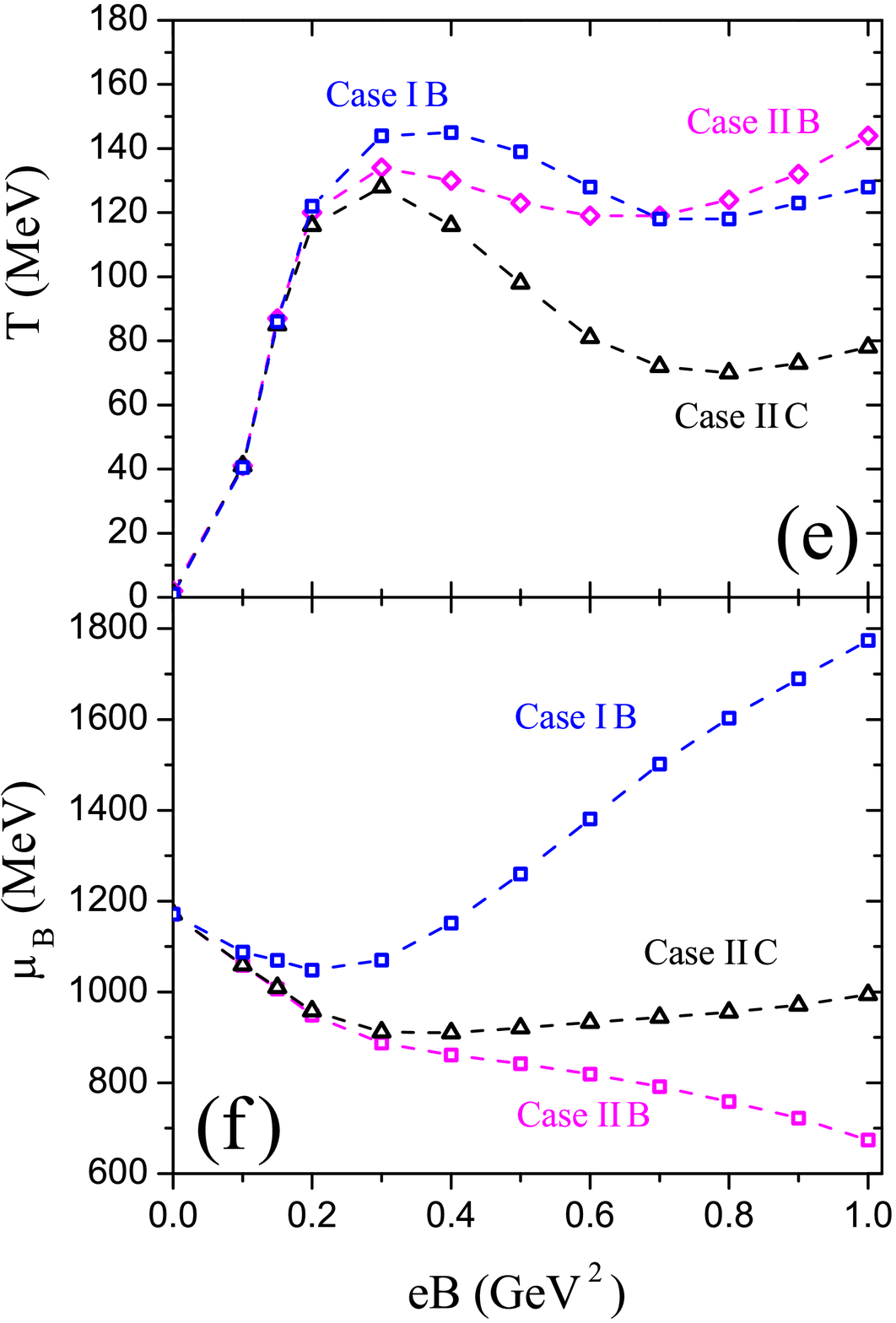}
    \caption{The location of the CEP: (a) and (d) $T$ versus $\mu_B$, for different 
            intensities of the magnetic field; (b) and (e) $T$ as a function of $eB$; 
            (c) and (f) $\mu_B$ versus $eB$, for the vector interaction $\alpha=0.25$ 
            [left, (a), (b) and (c) panels] and $\alpha=0.71$ 
            [right, (d), (e) and (f) panels]. 
            The red and blue curves were obtained with no IMC effects and 
            correspond to Cases IA and IB. 
            The IMC effects and vector interaction were included in the black curves 
            ($G_V$ is fixed) and magenta ($G_V$ weakens with an increasing $eB$). 
            For more details see the text at the end of Sec. \ref{sec:IMC_CEP}.
            }
\label{fig:CEP_IMC_GV}
\end{figure*}

\subsection{$G_V\ne 0$}
\label{subsec:IMC_CEPb}

In this subsection we discuss the role of the vector interaction in the location 
of the CEP of magnetized quark matter taking into account explicitly the
inverse magnetic catalysis at finite $T$ through the renormalization of the 
coupling $G_s$ due to the presence of the magnetic field.
We will consider  two scenarios, with or without an explicit dependence of 
$G_V$ on the magnetic field: $G_V=\alpha G_s(eB)$ (Case IIB), and 
$G_V =\alpha G^0_s$ (Case IIC).
For the constant $\alpha$ we will take a general value $\alpha=0.25$ 
(left panels of Fig. \ref{fig:CEP_IMC_GV}),
and the critical value discussed above, $\alpha=0.71$ 
(right panels of Fig. \ref{fig:CEP_IMC_GV}) .

We will first consider $\alpha=0.25$ [see panels (a), (b) and (c) of 
Fig. \ref{fig:CEP_IMC_GV}]. 
In these panels, the red curves correspond to Case IB with $G_V=0.25 G_s^0$, 
already presented  in Fig. \ref{fig:CEP_eB_GV_2}, and the black and the blue 
curves are for $G_s=G_s(eB)$ and, respectively, for $G_V=0.25 G_s(eB)$ (Case IIB) 
and $G_V=0.25 G_s^0$ (Case IIC).
When the IMC effects are included in the definition of the scalar
coupling, and the vector term is taken into account, it is seen that
for small values of $B$ the CEP moves to lower values of $T$ and slightly higher 
values of $\mu_B$ (although a bit lower than in Case IIA), in both Cases IIB and 
IIC [see panels \ref{fig:CEP_IMC_GV}(b) and \ref{fig:CEP_IMC_GV}(c), respectively]. 
Although the coupling $G_s(eB)$, and consequently also $G_V$ in Case IIB, is 
slightly affected by the magnetic field,  the overall balance between the 
contributions of the attractive and the repulsive interactions is almost unchanged. 

For $0.09<eB\lesssim0.3$GeV$^2$  the behavior for all three cases with 
$G_V\ne 0$, IB, IIB, and IIC, is very similar to the one found with $G_V=0$ 
(Cases IA and IIA): the critical temperature increases 
[see panel \ref{fig:CEP_IMC_GV}(b)] and the critical chemical potential decreases 
[see panel \ref{fig:CEP_IMC_GV}(c)].
  
However, differences occur for stronger magnetic fields.
When $eB\gtrsim 0.3$GeV$^2$, the CEP moves to smaller chemical potentials while 
the temperature does not change much in Case IIB. 
On the other hand, in Case IIC, with a coupling $G_V$ that does not change with 
$B$, the vector contribution becomes more important than the catalysis effect 
of the magnetic field, since this last effect has been weakened by a weak 
coupling $G_s(eB)$. 
Consequently, as soon as $G_s$ is sufficiently weak, the effect of $G_V$ is seen 
in the decrease of $T^\text{CEP}$ [see panel \ref{fig:CEP_IMC_GV}(b)], while 
$\mu_B^\text{CEP}$ practically does not change [see panel \ref{fig:CEP_IMC_GV}(c)].

When $\alpha=0.71$ [see panels (d), (e) and (f) of Fig. \ref{fig:CEP_IMC_GV}] the CEP, 
for all cases with $G_V\ne 0$, shows a similar behavior as the field is increased 
from $eB=0$, when $T^\text{CEP}=0$ MeV and $eB\sim0.3$~GeV$^2$: the CEP moves to lower 
chemical potentials [see panel \ref{fig:CEP_IMC_GV}(f)] and the critical temperature 
to larger values [see panel \ref{fig:CEP_IMC_GV}(e)]. 
When the IMC effect is included, Cases IIB and IIC, the curves overlap due to a 
weakened coupling $G_s(eB)$. 
Above $eB\sim0.3$~GeV$^2$, in Case IIB the CEP occurs for smaller values of 
$\mu_B^\text{CEP}$ while $T^\text{CEP}$ does not change much, a behavior also occurring  
for  $\alpha=0.25$. In Case IIC, with  $G_V=0.71G_s^0$, the trend is 
different from the corresponding one obtained with $\alpha=0.25$: the
coupling $G_s$ becomes sufficiently weak with respect to the magnitude of $G_V$
and the effect of the vector coupling is seen in the decrease of $T^\text{CEP}$ and
the weak increase of $\mu_B^\text{CEP}$. For $eB>0.8$ GeV$^2$ the effect of the 
magnetic field clearly overlaps the effect of $G_V$, and the CEP goes to higher 
temperatures and chemical potentials as in  Case IB without the IMC effect 
($G_V=0.71G_s^0$).

\begin{figure}[t!]
\centering
  \includegraphics[width=1\linewidth,angle=0]{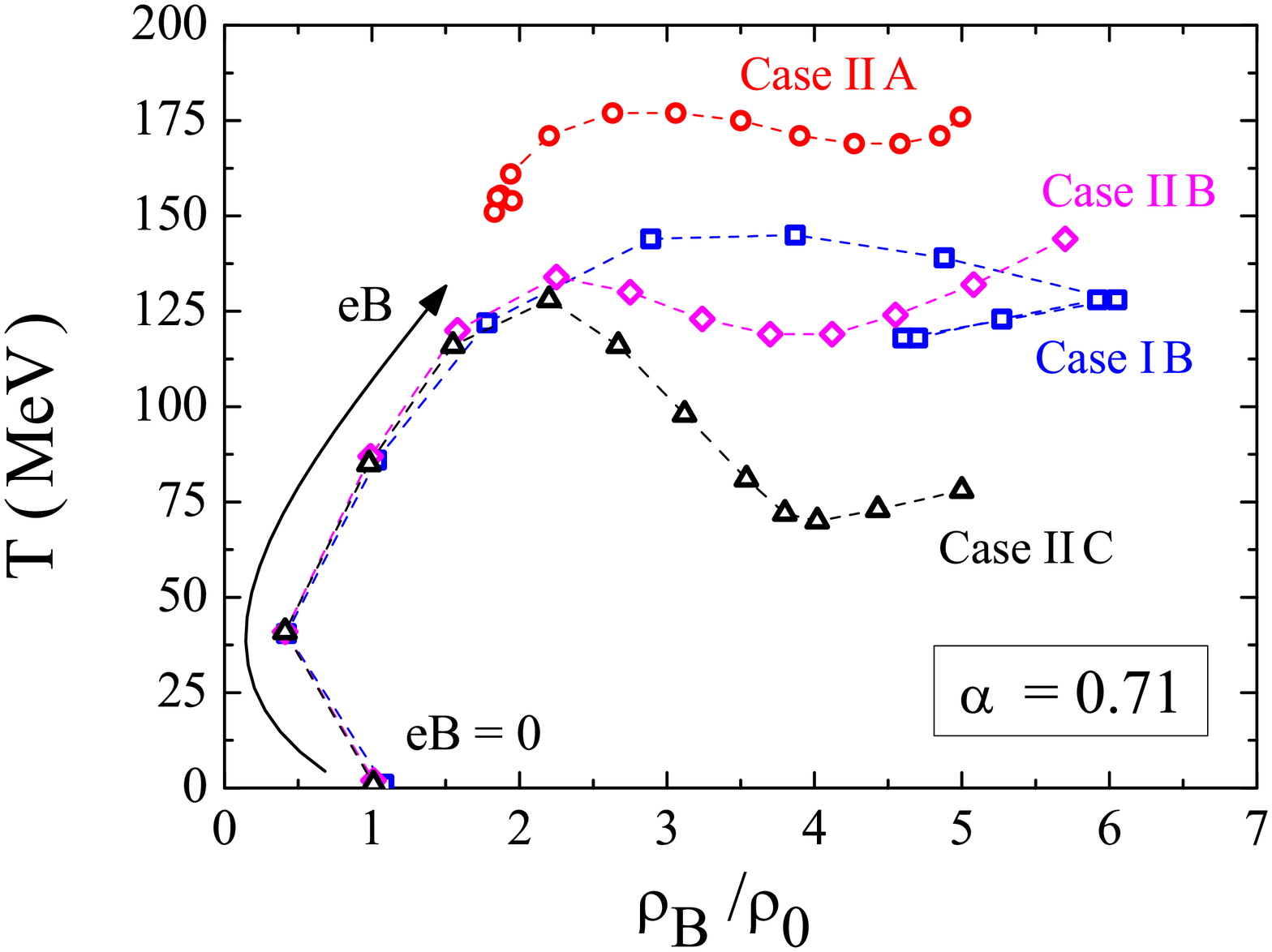}
  \caption{The CEP location on the T-$\rho_B$ plane. Cases IIA, IIB and
           IIC with IMC effects are compared with Case IB without IMC effects. 
           Case IIA is the only scenario excluding the vector interaction.
          }
\label{fig:CEP_IMC_GV_dens}
\end{figure}

It is also interesting to study the effect of the magnetic field on
the CEP baryonic density. In  Fig. \ref{fig:CEP_IMC_GV_dens} the CEP
location is plotted as a function of this quantity for different scenarios 
discussed in the present and previous sections. 
It is seen that the CEP baryonic density is not much affected 
by the field, the only exception if the different behavior of Case IB
for the stronger fields, the CEP temperature being the property that 
distinguishes the different scenarios.

\section{Conclusions}
\label{Conclusions}

In the present work we have discussed the possible consequences of the IMC 
effect on the location of the CEP.  The discussion has been performed within the  
(2+1) PNJL model with the possible inclusion of the vector interaction. 
Within this model the IMC effect is described by considering a scalar
coupling that weakens with the increase of the magnetic field as
proposed in \cite{Ferreira:2014kpa}. The dependence of the coupling on
the magnetic field has been fitted to the LQCD calculations of the transition
temperature at zero chemical potential, which show a decreasing
crossover temperature with an increase of the magnetic field for
intensities below $eB=1$ GeV$^2$.

The main conclusion of the present work  is that the IMC effect will have 
noticeable effects on the  location of the CEP. If the vector interaction is 
switched off, the CEP will occur at increasingly smaller chemical potentials 
and at a practically unchanged temperature if the field satisfies 
$eB\gtrsim 0.3$~GeV$^2$. 
This behavior is contrary to the findings of \cite{Avancini:2012ee} with SU(3) 
NJL with constant couplings, where it was shown that above $0.3$~GeV$^2$, both 
$T^\text{CEP}$ and $\mu_B^\text{CEP}$ increase.  
Also the baryonic density at the CEP is affected: 
including IMC effects it increases  only $~1/3$ of the expected if  IMC effects 
were not considered, making the CEP much more accessible in the laboratory. 
However, for weaker fields, $eB<0.3$~GeV$^2$, both scenarios give similar results.

If the vector interaction is included, we must consider two scenarios:
a strong enough vector interaction will turn the first order 
deconfinement/chiral transition into a crossover \cite{Fukushima:2008wg}  
or, on the contrary, the vector interaction is not strong enough to wash out the 
CEP, but moves its location to smaller temperatures and baryonic densities and 
to larger chemical potentials.
For the first scenario we have confirmed within the (2+1) PNJL model the 
findings of \cite{Denke:2013gha} obtained with the SU(2) NJL model, showing that 
a strong magnetic field would transform the crossover  into a first order phase
transition. With respect to the second scenario, we have shown that for
sufficiently small fields, the repulsive effect of the interaction is stronger 
than the MC effect originated by the magnetic field, and the CEP location occurs 
at smaller temperatures and slightly larger chemical potentials. The decrease of 
the CEP temperature could be quite large, for $\alpha=0.25$: $T^\text{CEP}$ suffers a 
reduction above 60 MeV when $eB$ goes from 0 to  0.09~GeV$^2$. Both effects 
corresponding to the two scenarios move the CEP to regions of temperature and 
density in the phase diagram that could be more easily accessible to HIC. 

For larger fields,  0.09~GeV$^2< eB \lesssim 0.3$ GeV$^2$ the magnetic field 
wins and $\mu_B^\text{CEP}$ decreases while $T^\text{CEP}$ increases, showing a behavior 
similar to the corresponding one in the absence of a vector interaction.
Above $ eB \gtrsim 0.3$ GeV$^2$ the CEP chemical potential increases but the CEP 
temperature keeps practically unchanged. 

The joint effect of the vector interaction and the IMC effect  of the magnetic 
field depends strongly on the magnitude of the magnetic field and whether
the vector coupling  becomes weaker with the magnetic field. We have considered 
two scenarios: a constant $G_V$ and a $G_V$ that weakens when the magnetic 
field intensity increases. In the second case the location of the CEP for  very 
strong magnetic fields is not much affected by the vector contribution, the 
magnetic field defining the structure of the phase transition. 
If, however, the vector coupling is not affected by the magnetic field, the 
weakening of the scalar coupling with the increase of the magnetic field 
intensity leads to the dominance of the vector contribution, which translates 
into a reduction of the CEP temperature.  An overall general conclusion is
the reduction of the CEP temperature when the vector interaction is included. 
This effect will be quite strong if the vector coupling does not become weaker 
when the magnetic field intensity increases.

\vspace{0.25cm}
{\bf Acknowledgments}:
This work was supported by ``Fundação para a Ciência e Tecnologia", Portugal, 
under the Grants No. SFRH/BPD/102273/2014 (P. C.), No. SFRH/BD/51717/2011 (M. F.),
and No. SFRH/BPD/63070/2009 (J. M.). 
This work was partly supported by Project PEst-OE/FIS/UI0405/2014 developed 
under the initiative QREN financed by the UE/FEDER through the program
COMPETE-``Programa Operacional Factores de Competitividade'', by 
"Conselho Nacional de Desenvolvimento Científico e Tecnológico" (CNPq), Brazil, 
and by ``NewCompstar'', COST Action MP1304.

\end{document}